\documentclass[a4paper,fleqn,usenatbib]{mnras}

\usepackage{mathptmx}
\usepackage[T1]{fontenc}
\usepackage{ae,aecompl}

\usepackage{graphicx}	% Including figure files
\usepackage{amsmath}	% Advanced maths commands
\usepackage{amssymb}	% Extra maths symbols
\usepackage{xspace}
\renewcommand\ion[2]{#1$\;${\small\rmfamily \uppercase\expandafter{\romannumeral #2 \relax}}}

\newcommand{\halpha}{H\hspace{0.25pt}$\alpha$\xspace}
\newcommand{\hbeta}{H$\beta$\xspace}

\title[Ionized gas kinematics]{The MUSE Atlas of Disks (MAD): Ionized gas kinematic maps and an application to Diffuse Ionized Gas.}

\author[M. den Brok et al.]{
  Mark den Brok,$^{1,2}$\thanks{E-mail: mdbrok@aip.de} C. Marcella Carollo,$^{1}$ Santiago Erroz-Ferrer,$^{1}$ Martina Fagioli,$^{3}$
  \newauthor Jarle Brinchmann,$^{4,5}$ Eric Emsellem,$^{6,7}$ Davor Krajnovi\'c,$^{2}$ Raffaella A. Marino,$^1$ 
  \newauthor Masato Onodera,$^{1,8}$ Sandro Tacchella,$^{1,9}$ Peter M. Weilbacher$^2$ and  Joanna Woo$^{1,10}$ 
\\
$^{1}$Department of Physics, ETH Z\"urich, Wolfgang-Pauli-Str 27, 8093, Z\"urich, Switzerland\\
$^{2}$Leibniz-Institut f\"ur Astrophysik Potsdam (AIP), An der Sternwarte 16, 14482 Potsdam, Germany\\
$^{3}$Institute for Particle Physics and Astrophysics, ETH Z\"urich, Wolfgang-Pauli-Str 27, 8093, Z\"urich, Switzerland\\
$^{4}$Instituto de Astrof{\'\i}sica e Ci{\^e}ncias do Espa{\c c}o, Universidade do Porto, CAUP, Rua das Estrelas, PT4150-762 Porto, Portugal\\ 
$^{5}$Leiden Observatory, Leiden University, PO Box 9513, NL-2300 RA Leiden, the Netherlands\\
$^{6}$European Southern Observatory, Karl-Schwarzchild-Str. 2, Garching bei M\"unchen, 85748, Germany\\
$^{7}$Univ Lyon, Univ Lyon1, ENS de Lyon, CNRS, Centre de Recherche Astrophysique de Lyon UMR5574, F-69230 Saint-Genis-Laval France \\
$^{8}$Subaru Telescope, National Astronomical Observatory of Japan, HI 96720 Hilo, USA\\
$^{9}$Harvard-Smithsonian Center for Astrophysics, 60 Garden St., Cambridge, MA 02138, USA\\
$^{10}$Department of Physics \& Astronomy, PO Box 1700 STN CSC, Victoria BC V8W 2Y2, Canada\\
}

\date{Accepted XXX. Received YYY; in original form ZZZ}

\pubyear{2019}

\begin{document}
\label{firstpage}
\pagerange{\pageref{firstpage}--\pageref{lastpage}}
\maketitle

\begin{abstract}
  We have obtained data for 41 star forming galaxies in the MUSE Atlas of Disks (MAD) survey with VLT/MUSE. These data allow us, at high resolution of a few 100 pc, to extract ionized gas kinematics ($V, \sigma$) of the centers of nearby star forming galaxies spanning 3 dex in stellar mass. This paper outlines the methodology for measuring the ionized gas kinematics, which we will use in subsequent papers of this survey.  We also show how the maps can be used to study the kinematics of diffuse
  ionized gas for galaxies of various inclinations and masses. Using two
  different methods to identify the diffuse ionized gas,  we measure rotation
  velocities of this gas for a subsample of 6 galaxies. We find that the
  diffuse ionized gas rotates on average slower than the star forming gas with
  lags of  0-10 km/s while also having higher velocity dispersion. The
  magnitude of these lags is on average 5 km/s lower than observed
    velocity lags between ionized and molecular gas. Using Jeans models to
  interpret the lags in rotation velocity and the increase in velocity
  dispersion we show that most of the diffuse ionized gas kinematics are
  consistent with its emission originating from a somewhat thicker layer than
  the star forming gas, with a scale height that is lower than that of
  the stellar disk.
%which does not extend beyond the stellar disk.

\end{abstract}

\begin{keywords}
%keyword1 -- keyword2 -- keyword3
galaxies: spiral -- galaxies: kinematics and dynamics
\end{keywords}

%%%%%%%%%%%%%%%%%%%%%%%%%%%%%%%%%%%%%%%%%%%%%%%%%%

%%%%%%%%%%%%%%%%% BODY OF PAPER %%%%%%%%%%%%%%%%%%

\section{Introduction}
The global motions of stars and gas in galaxies are determined primarily by the 
mass distribution inside the galaxy. In particular, the analysis of rotation curves in the outer parts of galaxies based on neutral hydrogen observation has led to the firm establishment of dark matter in galaxies \citep[e.g.][ and many others]{RubForTho80,Bos81,vanBahBeg85}. Although the 21 cm line has been one of the most widely used lines for this, it is also possible to trace rotation curves with ionized \citep[e.g.][]{MatForBuc92,GarMarAmr02,ErrKnaLea16} or molecular gas \citep[e.g.][]{Sof96,SofTutHon97}. As these emission lines are produced by gas in different physical states, they are often found in different parts of the galaxy. 

Much can be learned by comparison of kinematic tracers with each other. By comparing the velocity dispersion of low surface brightness CO, \citet{CalSchWal13} and \citet{MogdeBCal16} provide evidence for a faint, diffuse, higher dispersion CO component in nearby spiral galaxies that appears similar in dispersion (and therefore in thickness) as the neutral hydrogen. The kinematics of different tracers can be different. Molecular gas has on average a lower velocity dispersion than atomic gas. Gas dynamical tracers have in general a lower velocity dispersion than stars. For stars in spiral galaxies, the velocity dispersion is in turn a function of the age of the population, with older stars showing higher dispersion and a bigger vertical extent \citep{Wie77,CarDawHsu85}. 

The velocities of ionized gas should be very close to those of the molecular gas, as the stars that are responsible for the ionization have formed very recently from molecular gas and have a low asymmetric drift \citep[e.g.][]{QuiGuhChe18}. However, molecular and ionized gas do not always agree with each other. \citet{DavAlaBur13} study early-type galaxies and find that the ionized gas, contrary to the molecular gas, does not necessarily trace the circular velocity of the galaxies, mostly because of a different distribution of the molecular gas. Recently \citet{LevBolTeu18} compared $^{12}$CO (J=1--0) rotation curves with \halpha\ rotation curves derived from CALIFA data \citep{SanKenGil12}. They found that the \halpha\ gas usually, but not always, shows velocity lags, with median values between 0--25 km/s, with respect to the CO gas, which they attribute to the presence of extraplanar diffuse ionized gas.

First identified in the Milky Way (\citealt{Rey84}, but already suggested by \citealt{HoyEll63}), a major part of the \halpha\ emission from star forming galaxies comes from diffuse ionized gas (DIG) instead of directly from \ion{H}{2} regions \citep{Rey90,WalBra94,ZurRozBec01}. The excitation mechanisms of this diffuse gas are not completely understood, but have been attributed to UV radiation from leaky \ion{H}{2} regions \citep{HoyEll63,Rey84}, cosmic rays \citep[e.g.][]{DahDetHum94,VanWooGir18}, turbulent mixing layers \citep{HafDetBec09,BinFloRag09} and evolved stars \citep[e.g.][]{KapJogKew16,ZhaYanBun17}. All of these lead to different line ratios in optical strong emission lines \citep{HafReyTuf99,HooWal03,MadReyHaf06}. The layer of ionized gas in the Milky Way, the Reynolds layer, extends to beyond a kpc above the galactic disk \citep{Rey89}. Also in external galaxies, such diffuse gas layers have been found \citep[e.g.]{RosDet03}, extending on average 1-2 kpc above the galactic plane.

The study of the kinematics of this DIG is interesting as it potentially could reveal the origin and ionization source of the DIG.

The kinematics of ionized gas have traditionally been studied with long slits \citep[e.g.][]{RubForTho80,MatForBuc92}, as well as Fabry-Perot (FP) interferometers  \citep{Tul74,RydZasMcI98,HerFatCar08}. Recently, 2D mapping of ionized gas has become possible by the developments of integral field units such as SAURON \citep{BacCopMon01}, SparsePak \citep{BerAndHar04}, PMAS \citep{RotKelFec05} and VIRUS-P \citep{HilMacSmi08} to study ionized gas kinematics in late-type galaxies \citep{GanFalPel06,BerVerWes10,GarMarBar15}.

Each of these methods for studying ionized gas has its advantages and disadvantages. The FP interferometers provide higher spectral resolution but at the cost of wavelength baseline. Some of the mentioned IFUs have lower spectral resolution and/or lower spatial resolution than the FP interferometers, but provide a longer wavelength range. The Multi Unit Spectroscopic Explorer (MUSE) instrument \citep{BacAccAdj10} on the Very Large Telescope (VLT) falls in between the properties of the mentioned instruments. Although its spectral resolution is not as high as that of FP interferometers, its spatial resolution is superb, and its large spectral range at intermediate resolution, combined with a 1\arcmin $\times$ 1\arcmin\ field of view, makes this instrument also suitable to study more diffuse gas that has surface brightness comparable to the stellar continuum emission. This is a particular niche for MUSE that has been difficult to observe with previous instruments.

In this paper we compare the kinematics of the DIG with that of the star
forming gas for a sample of star forming galaxies with various inclinations. Previous studies of the kinematics of the diffuse ionized gas have mainly focused on extraplanar gas in (almost) edge-on galaxies such as NGC 891 \citep{HeaRanBen06}, NGC 4302 \citep{HeaRanBen07},  NGC 5775 \citep{TulDetSoi00,Ran00,HeaRanBen06b} and NGC 2403 \citep{FraOosSan04}, and recently on survey-scale with 67 edge-on galaxies in the Mapping Nearby Galaxies at APO survey \citep[MaNGA,][]{BizWalYoa17}. There seems to be a consensus that the rotational velocity of the gas above the plane is lower than that of the gas in the midplane, although this gas may show evidence for non-circular motions \citep{FraOosSan04}. There exist no measurements of the kinematics of diffuse gas in relatively face-on galaxies with the exception of the work of \citet{BoeGalZwe17}, who find also lags in the rotation velocity of DIG in M 83.

For the extra-planar gas,  models have been developed to explain this slower rotation. \citet{ColBenRan02} use ballistic models to explain this gas.  The clouds, which are launched from the midplane of the galaxy disk, show a decrease in rotation signal as they travel away from the plane. The clouds also move radially outward due to a combination of conservation of angular momentum and lower gravitational force. \citet{BarCioFra06} \citep[see also][]{Ben02} proposed hydrostatic models to predict the rotation velocity of gas outside the galactic midplane. These models also explain the velocities of extraplanar diffuse gas, although they require high temperatures ($10^4 \lesssim T \lesssim 10^6$ K) for the gas, which might be more appropriate for galactic coronae than for the cold gas that has been observed. \citet{FraBin06} and  \citet{MarFraCio10}  develop models in which there is a continuous launching of clouds from the disk, which eventually lose velocity due to a drag with a hot corona. 

This paper presents some of the first results of the MUSE Atlas of Disks (MAD) Survey. This survey is mapping out the inner parts of a sample of  45 (mostly) nearby star forming galaxies with MUSE. The goal of the survey is to understand the formation and evolution of disc galaxies through studies of the kinematics of gas and stars in these galaxies, together with their star formation histories to understand how disk galaxies have formed and evolved. The description of the survey will be presented in Carollo et al. (in prep., henceforth Paper 1), as well as in \citet[][henceforth Paper 2]{ErrCarden19} which describes the resolved metallicity and star formation properties in the galaxies. With the high spatial resolution of MUSE, we check the scenario proposed by \citet{LevBolTeu18} that \halpha\ rotation curves may be lagging the  $^{12}$CO (J=1--0) rotation curves because of the presence of extraplanar DIG. We do this by separating the star forming gas from the diffuse ionized gas and analyzing the kinematics separately. We also outline the procedures used for deriving the kinematics, which we will use in future papers. This paper is structured as follows. In Section \ref{sec:data}, we briefly recapitulate the sample selection and data reduction. In Section \ref{sec:kin} we present the procedures for the derivation of the kinematics and present the kinematic maps. Sec. \ref{sec:dig} shows the methods to identify DIG and the rotational velocity measurements. The rotation velocities of DIG and star forming (SF) gas are then presented in Sec. \ref{sec:results}, followed by a discussion in Sec. \ref{sec:discussion}.

\section{Observations and data reduction}\label{sec:data}
The MAD survey has observed a sample of $45$ nearby star forming galaxies with VLT/MUSE as part of a Large Guaranteed Time Program and provides us with high spatial and spectral resolution observations of the centers of these galaxies. Although the sample selection and data reduction will be described in detail in paper I (Carollo et al. in prep.), we briefly summarize the selection criteria and reduction procedures here as well.

\begin{figure*}
 \includegraphics[width=1.02\textwidth]{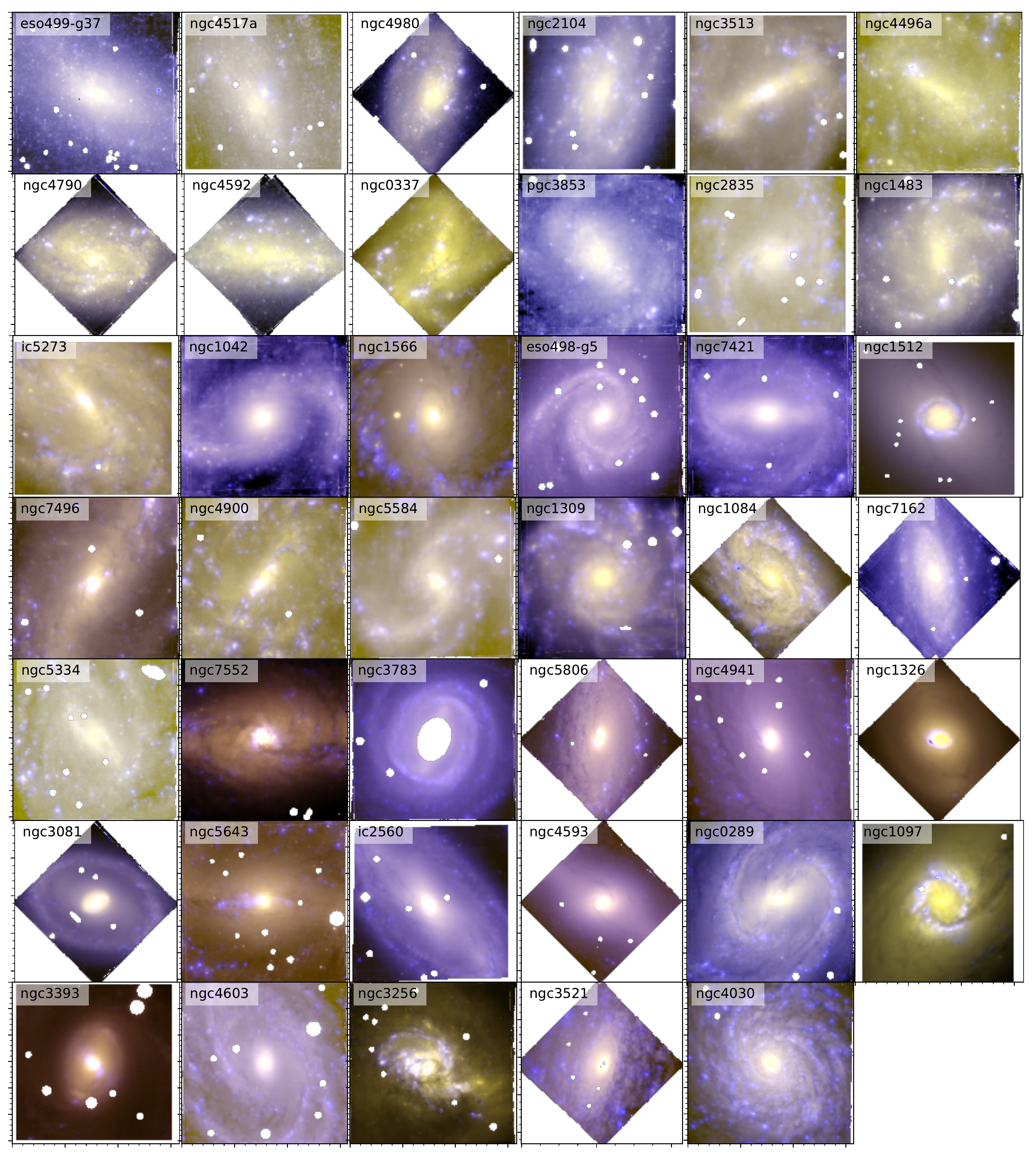}
 \caption{False colour images of the galaxies in the sample, based on the
   white light maps (collapsed cube) of the galaxies with in blue the ionized
   gas as traced by \halpha.}
 \label{fig:2dmaps_all_gals}
\end{figure*}

Besides observability constraints (visible for at least one consecutive hour
per night from VLT with airmass $< 1.5$), galaxies were selected to be bright
(M$_B < 13$) and on the star forming main sequence
\citep{BriChaWhi04,NoeWeiFab07,DadDicMor07}. In order to make optimal use of
the field of view of MUSE, a size cut was applied to ensure that galaxies were
not significantly smaller than the field of view and that our observations
would extend to at least 0.75 effective radius. To ensure a uniform coverage
of spectral features, the host galaxy redshifts were limited to the range
z=0.002-0.012. Target galaxies were additionally selected to have good-quality
optical Hubble Space Telescope (HST) imaging in at least one red passband
(F606W, F814W or equivalent) available from either HST/WFPC2, HST/ACSWFC or
HST/WFC3, moderate inclinations ($0.3 < \epsilon < 0.95$), and limited
foreground extinction. The final hand-picked sample contains 45 galaxies. In
this paper we discuss 41 sample galaxies; we note that the remaining 4
galaxies were chosen as a reference sample containing merging galaxies. The
data reduction and analysis of these remaining galaxies is more complicated,
as two of these galaxies are at cosmological distances, and the other two are
large mosaics. As the kinematics of these particular 4 galaxies have little
relevance for this paper, we postpone their presentation to a future
paper. The 41 sample galaxies are listed in Table \ref{tab:sample} and shown
in Fig. \ref{fig:2dmaps_all_gals}. All
galaxies were observed as part of the MUSE GTO observations, except for NGC
337, which was observed during the commissioning of MUSE, and NGC 1097, which
was observed by program 097.B-0640 (PI Gadotti).

\begin{table}
\caption{Properties of the MAD sample. Distances were obtained from NED. Effective radii are based on 2D decompositions of 2Mass data. Stellar masses were derived from optical and NIR SED fitting (paper I). }
\label{tab:sample}  
\begin{tabular}{lccc}
\hline
Name  & Distance[Mpc] & R$_e$  [\arcsec]     & log(M$_{\star}$/M$_{\odot}$) \\
\hline
ESO 499-G37  & 18.3 &  18.3  & 8.5 \\
  NGC 4517A  &  8.7  &  46.8 &  8.5  \\
  NGC 4980    & 16.8 &  13.0 &  9.0  \\
  NGC 2104     &  18.0  & 16.5 &  9.2  \\
  NGC 3513   &  7.8  &  55.4 &  9.4  \\
  NGC 4496A & 14.7 & 37.1 & 9.5 \\
  NGC 4790    &   16.9 &  17.7 &  9.6   \\
  NGC 4592   &   11.7 &  37.9 &  9.7  \\
  NGC  337   &  18.9 &  24.6 &  9.8  \\
  PGC 003853     &  11.3 &  73.1  & 9.8  \\
  NGC 2835     &   8.8  &  57.4  & 9.8   \\
  NGC 1483    &  24.4 &  19.0  & 9.8  \\
  IC 5273    &  15.6 & 33.8 & 9.8 \\ 
  NGC 1042     &  15.0 &  63.7  & 9.8  \\
  NGC 1566     &  6.6 & 60.3   & 9.9 \\
  ESO 498-G5   &  32.8 &  19.8  & 10.0 \\
  NGC 7421     &  25.4 &  29.6  & 10.1 \\
  NGC 1512     & 12.0 & 63.3   & 10.2 \\
  NGC 7496   &  11.9 &  66.6   & 10.2 \\
  NGC 4900    &  21.6 &  35.4  & 10.2 \\
  NGC 5584     &  22.5 &  63.5  & 10.3 \\
  NGC 1309    &  31.2 &  20.3  & 10.4 \\
  NGC 1084    &  20.9 &  23.8  & 10.4 \\ 
  NGC 7162    &  38.5 &  18.0  & 10.4 \\
  NGC 5334    &  32.2 &  51.2  & 10.6 \\
  NGC 7552   &  22.5 &  26.0  & 10.6 \\
  NGC 3783    &  40.0 &  27.7  & 10.6 \\
  NGC 5806    &  26.8 &  27.2  & 10.7  \\
  NGC 4941     &  15.2 &  64.7  & 10.8  \\
  NGC 1326    &  18.9 &  26.2  &  10.8 \\   
  NGC 3081   &  33.4 &  18.9  & 10.8 \\
  NGC 5643     &  17.4 &  60.7  & 10.8 \\
  IC 2560      &  32.2 &  37.7  & 10.9 \\
  NGC 4593   &  25.6 &  63.3  & 11.0 \\
  NGC 289    &  24.8 &  27.0  & 11.0  \\
  NGC 1097    &  16.0 &  55.0   & 11.1 \\ 
  NGC 3393    &  55.2 &  21.1  & 11.1 \\
  NGC 4603     &  32.8 &  44.7  & 11.1  \\
  NGC 3256     &  38.4 &  26.6  & 11.1 \\
  NGC 3521    &  14.2 &  61.7  & 11.2 \\
  NGC 4030    &  29.9 &  31.8   & 11.2 \\
 \hline
  \end{tabular}
  \end{table}

We reduced the data with the MUSE Instrument Pipeline \citep[][version 1.2 to 2.2, depending on when the galaxy was observed]{WeiStrPal16}. We used the pipeline to perform basic steps like bias subtraction, flat fielding and wavelength calibration. Each galaxy was observed with multiple (at least 3) exposures, which were interlaced with sky integrations. The total on-target exposure time was always 1 hour per galaxy. We experimented with different sky observation strategies for the first few galaxies, until we decided on a Object-Sky-Object observation pattern with 2 minute sky observations. To minimize the influence of bad pixels, the on-target observations were dithered with a small few-arcsecond dither pattern. We aligned the individual exposures by generating a narrowband \halpha\ image for each cube and using the image registration task \textsc{tweakreg} from the DrizzlePac package \citep{AviHacCar15} to measure shifts between different cubes. We then used the pipeline to drizzle the individual exposures to a common reference cube for each galaxy.

We used \textsc{zap} \citep{SotLilBac16} to subtract the sky from the individual exposures. \textsc{zap} performs a principal component analysis (PCA) on the separate sky observation, and encaptures the variation in the line spread function and the Poisson noise of bright sky lines in this way. After subtraction of the average sky determined on the sky frame from a target observation, \textsc{zap} also subtracts residuals in the sky subtraction by fitting the PCA components to the spectra. Our galaxies were processed with \textsc{zap} versions 1.0 and 2.0, depending on the observation date of the galaxy. We inspected the sky subtracted by \textsc{zap} to ensure that no galaxy light was subtracted.

The sky subtracted cubes were then combined to a single cube for each galaxy by taking the median value. We then corrected this cube for Milky Way foreground extinction using the values of \citet{SchFin11}, and masked foreground stars based on a by-eye identification on HST images and spectral identification in the MUSE cubes. We excluded the centre of NGC 3783 from the analysis as the broad Balmer lines from the type 1 AGN dominate most of the spectrum in this region. 

\section{Derivation of the kinematics}\label{sec:kin}

In order to derive accurate velocities and dispersions from gas emission lines, we first remove the stellar continuum by modeling this continuum with synthesized stellar templates. As the signal-to-noise of the continuum is not high enough to obtain good fits to the stellar continuum, we bin the data using the \textsc{voronoi} package of \citet{CapCop03}. To avoid stellar absorption lines and sky emission lines we use a 100 \AA\ wide area around 5700 \AA\ to determine the signal-to-noise. We then bin the spectra to achieve a S/N of 50 per \AA.

We fit PEGASE-HR templates \citep{LeBRocPru04} to the stellar continuum using \textsc{pPXF} \citep{Cap17}. The spectral resolution of this library (R=10000) is higher than the MUSE resolution and we therefore convolve the libraries with a wavelength dependent line spread function (LSF). We determine the MUSE LSF using both arc and sky lines, and after ensuring that our findings are similar to the one used by \citet{GueKraEpi17} we adopt their parametrization of the LSF. Before fitting with pPXF we mask all spectral pixels within a 400 km/s window of a strong emission line (Tab. \ref{tab:fitlines}) as well as the region around the Na D absorption lines. During the fit we allow for an additive polynomial up to order 4. 

For analyzing the ionized gas emission lines we subtract the best-fit continuum pixel-by-pixel. This is justified as long as the stellar populations do not change radically within one Voronoi bin. For each spaxel we scale the best-fit spectrum to the spectrum of the spaxel using the median value of each spectrum in the fitted range. We do not see evidence for any systematics in the continuum subtracted cubes. Further discussion on the reliability and possible influence of the continuum subtraction can be found in Sec. \ref{sec:contsub}.

We then proceed to analyze the emission lines. Even though the emission lines are often much brighter than the continuum, it is in many cases still necessary to bin pixels together to get a sufficiently high S/N for analyzing the emission lines kinematics. In order to estimate the signal-to-noise of the \halpha line without knowing its actual velocity, flux or dispersion we take the height of the \halpha line divided by the average dispersion of the surrounding continuum as a measure for the signal-to-noise. This is usually a lower limit to the significance with which the line is detected, but can sometimes lead to an overestimate of the S/N. After we have estimated the S/N per pixel, we tesselate the galaxy to reach S/N of 10 per bin using the aforementioned  \textsc{voronoi} code. For the most massive galaxies in the sample (e.g. NGC\ 4030, NGC\ 3521), this leads to bins with essentially the size of 1 spaxel.
\begin{table}
  \caption{Fitted emission lines.}\label{tab:fitlines}  
  \begin{tabular}{cc}
    \hline
    Transition & wavelength [\AA]\\
    \hline
    \hbeta & 4861.32 \\
    $[$\ion{O}{3}$]$  &   4958.91 \\
    $[$\ion{O}{3}$]$  &      5006.84 \\
    $[$\ion{O}{1}$]$    &    6300.30 \\
    $[$\ion{N}{2}$]$   &  6548.04 \\
    $[$\ion{N}{2}$]$   &  6583.46 \\
    \halpha &   6562.80 \\
    $[$\ion{S}{2}$]$   &  6716.44 \\
    $[$\ion{S}{2}$]$   & 6730.81 \\
    \hline
  \end{tabular}
  \end{table}

We fit a Gaussian line for each of the emission lines to the spectra binned using the \halpha\ tesselation. We note that some authors \citep[e.g.][]{BoeGalZwe17} use combinations of broad and narrow Gaussians, but given the 2.5 \AA\ spectral resolution of MUSE such a decompostion is not always warranted by our data. However, we will discuss the use of such a decomposition in Appendix \ref{sec:twocomp}. We broaden each line by the LSF at that wavelength and a velocity dispersion.

Different lines can trace gas with different physical conditions and therefore do not need to necessarily have the same kinematics. Here, we divide the lines in two groups, one containing the two visible Balmer lines, and the other group the other lines from Tab. \ref{tab:fitlines}. Lines inside each group share common kinematics ($V$, $\sigma$) but not common fluxes. Both groups of lines are fit simultaneously to the continuum-subtracted binned spectra using the Levenberg-Marquardt algorithm. Unless specified otherwise, the kinematics used in the remainder of the paper are based on the kinematics of the two Balmer lines.

\begin{figure*}
 \includegraphics[width=1.02\textwidth]{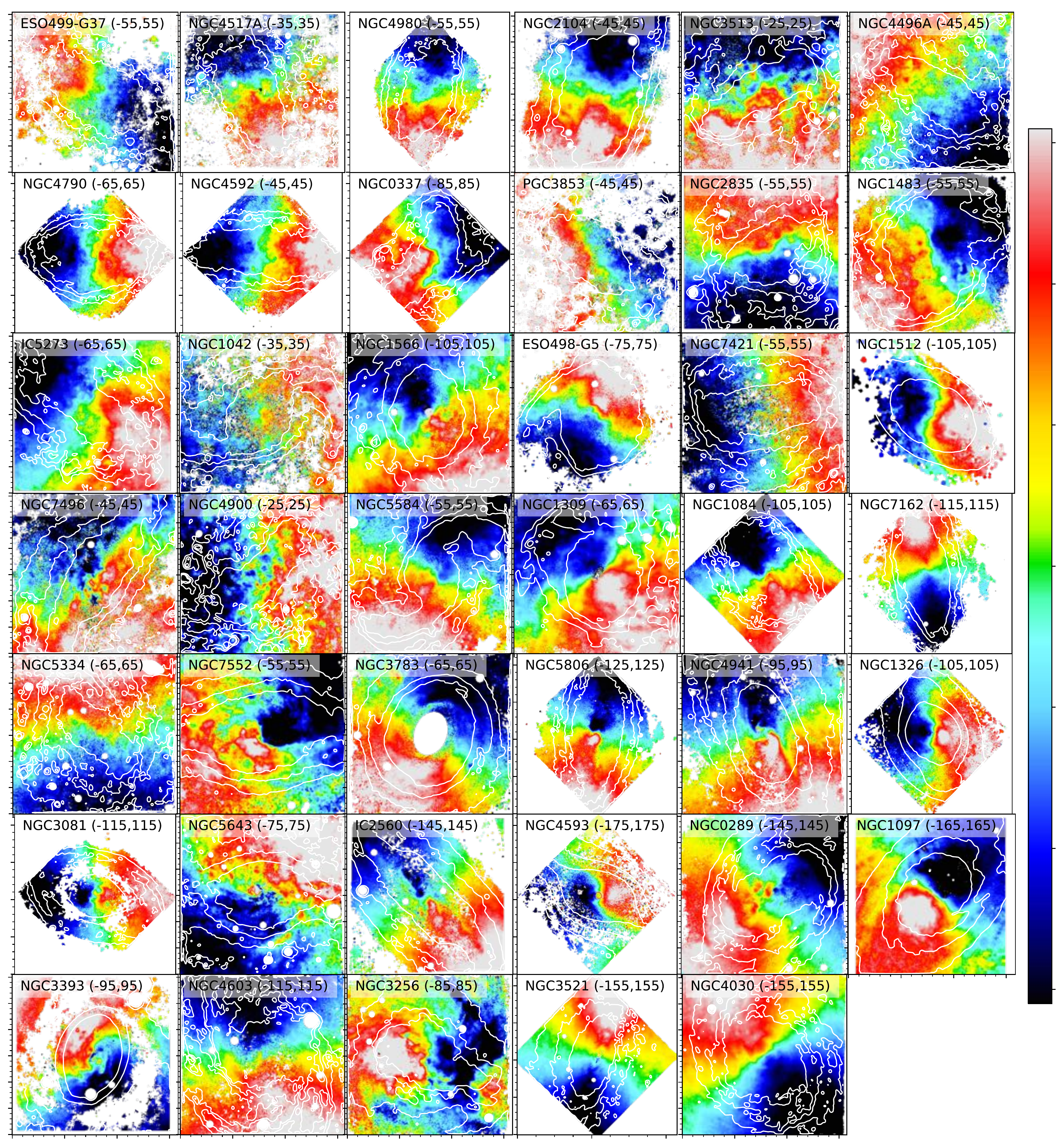}
 \caption{Maps of the velocity of the ionized gas derived from \halpha and \hbeta. White contours trace the shape of the galaxy as seen in the white light image and are logarithmically spaced in brightness between the 30\% and 95\% brightness levels. North points up in every map. All regular panels are $1$\arcmin$\times1$\arcmin; the rhombus-shaped panels are 1\farcm4$\times$1\farcm4. The colour scale is linear between the two numbers in brackets in km/s.}
 \label{fig:2dmaps_vel}
\end{figure*}

\begin{figure*}
 \includegraphics[width=1.02\textwidth]{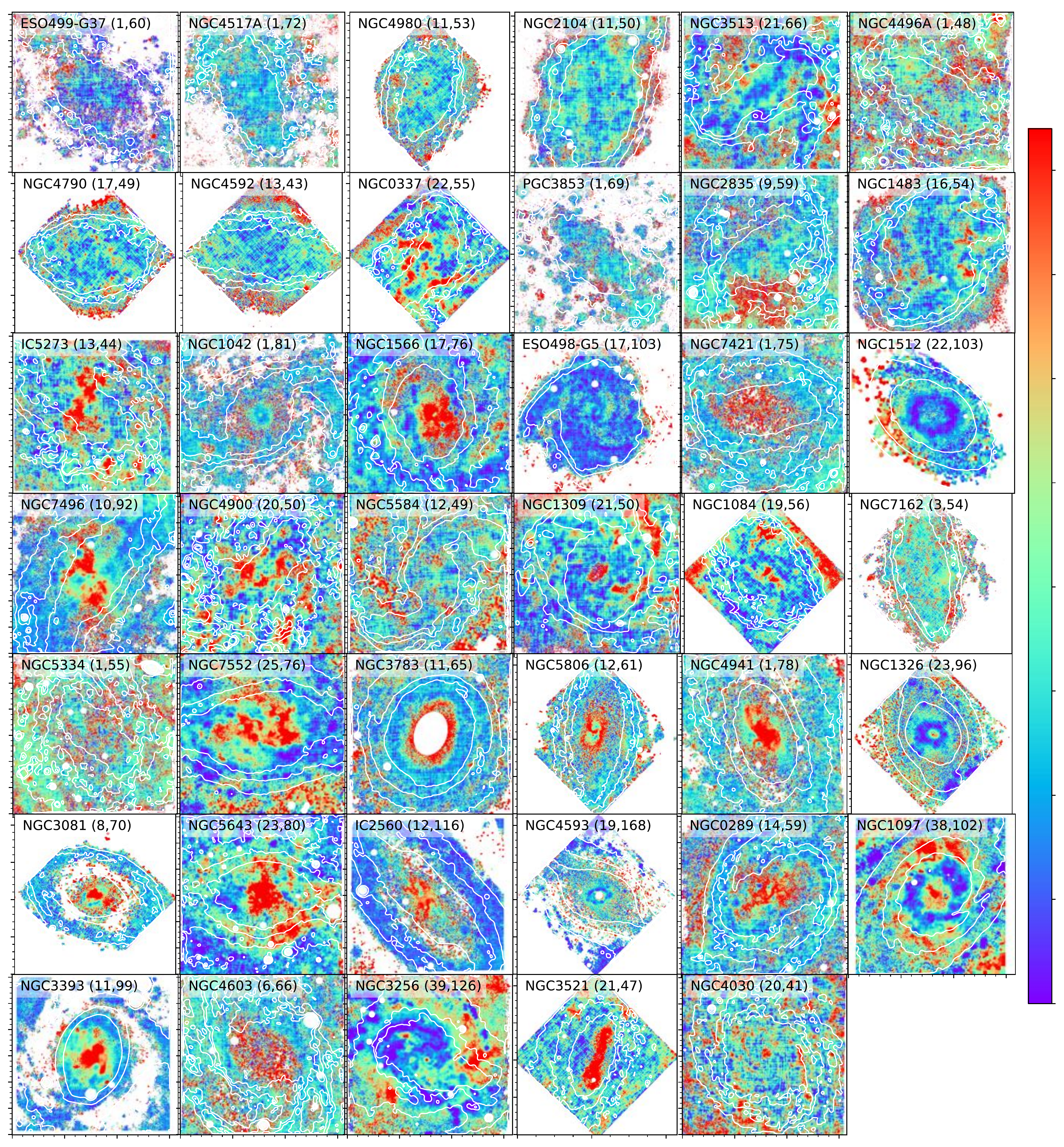}
 \caption{Maps of the dispersion of the ionized gas derived from \halpha and \hbeta. Contours and image sizes as in Fig. \ref{fig:2dmaps_vel}. The colour scale is linear between the two numbers in brackets in km/s.}
 \label{fig:2dmaps_sig}
\end{figure*}

We present the velocity and velocity dispersion of the \halpha and \hbeta lines in Figs. \ref{fig:2dmaps_vel} and  \ref{fig:2dmaps_sig}. The velocity maps show often irregular structure. NGC 3256 is a merging system \citep{Vor59} with two nuclei separated  $\sim 5$ arcsec in projection \citep{LirGonWar08} for which the observed velocity field is very irregular. NGC 337 is an asymmetric galaxy with an off-centred bar, which too has been suggested to be a merger \citep{SanBed94}.

We note that the velocity dispersions that we measure for the ionized gas are
often below the instrumental resolution. At the wavelength of \halpha, which,
out of the two Balmer lines in our spectral range, is the dominant line for
the determination of the kinematics because of its generally higher S/N, the
instrumental resolution is $\sigma\approx 47$ km/s. Velocity dispersions of
ionized gas in disk galaxies are often between 10-50 km/s
\citep[e.g.][]{EpiAmrMar08,ErrKnaLea15}. It is therefore a priori not clear
how reliable the determined velocity dispersions are. In Appendix
\ref{apx:sim_gas} we perform a set of idealized simulations to see if we can
recover the intrinsic dispersion of the gas. Our conclusions are that we see a
bias for dispersions below 25 km/s, and that, although maybe the absolute
values of the dispersions are biased, the relative values are still
robust. We note that the typical Voronoi bin size is much smaller than
  the scale over which the rotation velocity varies on order of the dispersion
  value, and that therefore beam smearing effects are not important.


\section{Kinematics of the DIG}\label{sec:dig}

\begin{table}
  \caption{Properties of the sub-sample used for deriving DIG velocities. Classification according to \citet{ButSheAth15}. Position angles were obtained from fits to the kinematic data. The axis ratios were taken from the photometric decomposition of S4G galaxies by \citet{SalLauLai15}. $f_0$ is the surface brightness cut-off value for DIG, defined in the text.}
  \label{tab:subsample}  
  \begin{tabular}{llccc}
    \hline
    Galaxy & Classification & P.A.  & q & f$_0$\\
           & & [deg] & & [$10^{-20}$ erg s$^{-1}$cm$^{-2}$]\\
    \hline
    NGC 4790 & Sm sp & -4 & 0.73 & 1580.2 \\
    NGC 4592 & SA(s)bc & 5 & 0.30  & 1329.9  \\
    NGC 1084 & SA(s)c & -54 & 0.54 & 1632.3 \\
    NGC 7162 & SAB(rs)bc & -80 & 0.42 & 476.0 \\
    NGC 3521 & SA(r'l,r)bc & 255 & 0.45 & 1606.3 \\
    NGC 4030 & SA(rs)bc & -54 & 0.78 & 1374.3 \\
    \hline
  \end{tabular}
  \end{table}

\subsection{Identification of the DIG}\label{sec:id_dig}
A critical step in the analysis is the identification of DIG. In edge-on galaxies, the identification is usually assumed to be the extra-planar gas. In less inclined galaxies, it is customary to either identify the star forming regions by the line ratios which differ from those for diffuse ionized gas, or by identifying peaks in the spatial distribution of emission line maps such as \halpha.

\begin{figure}
 \includegraphics[width=\columnwidth]{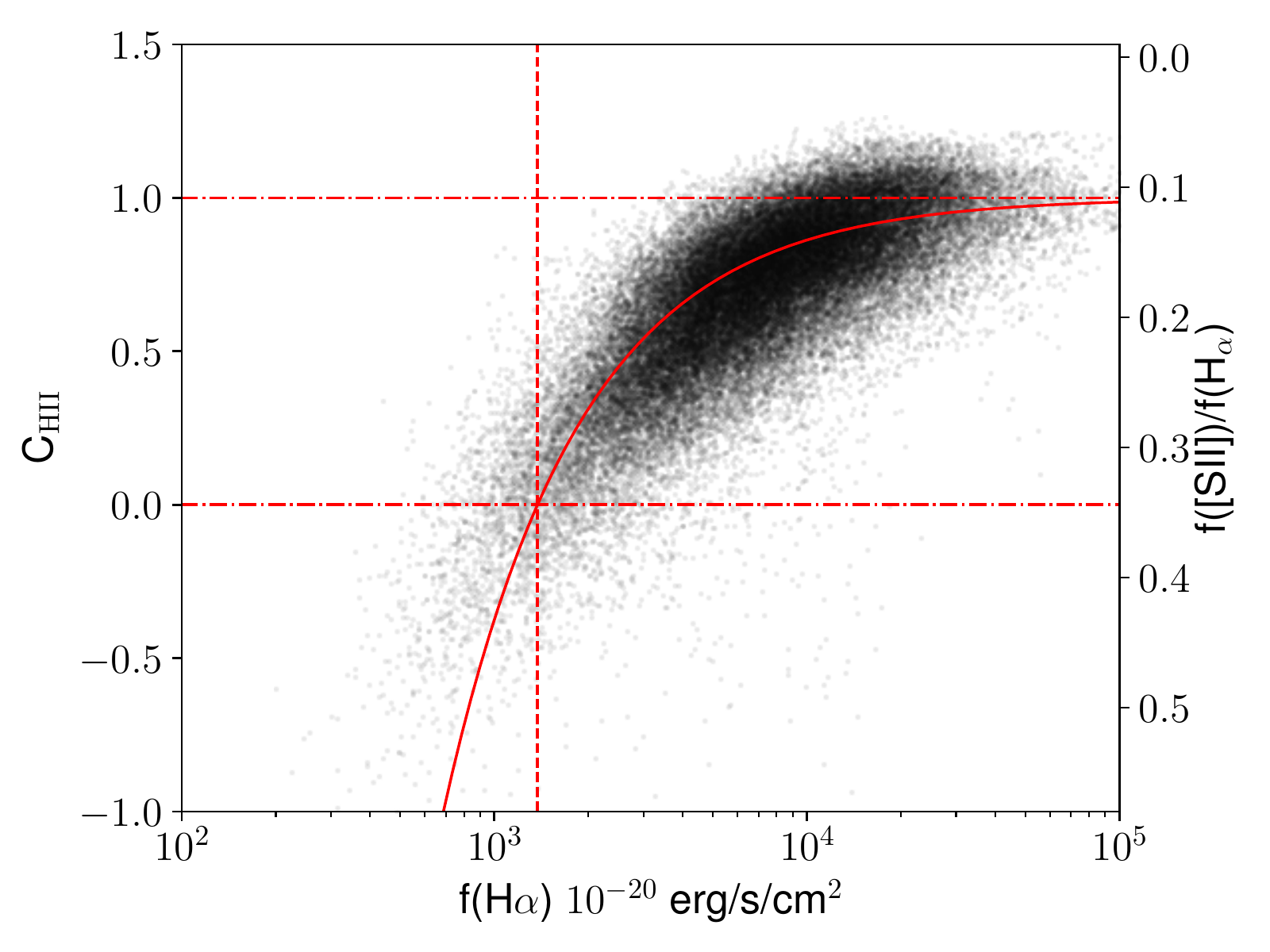}
 \caption{Ratio of [\ion{S}{2}]/\halpha\ for spaxels in NGC 4030. The dash-dotted lines show the fitted values for the [\ion{S}{2}]/\halpha\ ratio for pure \ion{H}{2} regions (upper line) and for diffuse gas (lower line). The vertical dashed line shows the location of $f_0$ (see text).}
 \label{fig:ngc4030_s2_ratio_halpha}
\end{figure}

In this paper, we use two methods to identify the DIG in our maps. The first method we use was developed by \citet{BlaHeiGeb09} and has since been used also by other authors to distinguish between DIG and star forming gas \citep[e.g.][]{KapJogKew16,KreBlaSch16}. This method assumes that the surface brightness of \halpha is composed of a contribution from DIG and of \ion{H}{2} regions. In regions where the surface brightness is below a cut-off, $f_0$, to be discussed below, it is assumed that the DIG is completely dominant and the DIG fraction $C_{DIG} = 1$, while at higher surface brightnesses we can write: $C_{DIG} = f_0/f_{H\alpha}$.  The surface brightness $f_0$, which determines this fraction, is based on measurements of the $[$\ion{S}{2}$]$/\halpha\ ratio throughout a galaxy, which is known to be much higher for diffuse ionized gas than star forming regions \citep{MadReyHaf06}. For our data we first correct the [\ion{S}{2}] and \halpha\ lines for extinction by using the Balmer decrement, assuming an intrinsic decrement of 2.86. Then we sum the fluxes of the two [\ion{S}{2}] 6716 and 6731 \AA\ lines together. We fit Formula 8 of \citet{BlaHeiGeb09} to the data using the MCMC code \texttt{emcee} \citep{ForHogLan13}, where we define the likelihood as the sum of the squared difference in the ratio divided by the squared errors on the ratio. Following Blanc et al., we adopt an intrinsic value of  $[$\ion{S}{2}$]$/\halpha\ = 0.35 for DIG and $[$\ion{S}{2}$]$/\halpha\ = 0.11 for \ion{H}{2} regions, both of which we scale with a free parameter $Z$ to account for metallicity differences between the MAD galaxies and the Milky Way (we refer the reader to the Appendix for results based on fits in which the metallicity was not free but adopted from Paper 2). To make the fit robust against outliers (see Fig. \ref{fig:ngc4030_s2_ratio_halpha}), we exclude the 5\% worst points from the likelihood. 

In Fig. \ref{fig:ngc4030_s2_ratio_halpha} we show for the galaxy NGC 4030 the
[\ion{S}{2}]/\halpha ratio as a function of \halpha surface brightness,
together with the fit shown in the solid red line. From this fit we decide how
to divide the spaxels in the datacube into those where, statistically,
\ion{H}{2} regions dominate ($f_{H\alpha} > 2 f_0$) and where DIG dominates
($f_{H\alpha} < 2 f_0$). The values of $f_0$ for the galaxies in the
  subsample are given in Table \ref{tab:subsample}. The value of $f_0$ found for NGC
  7162 is much lower than for the other galaxies. It is unclear if this
  because the Blanc criterion breaks down for lower spatial resolution or if
  the intrinsic properties of this galaxy are simply differrent.  

We also include a second identification method based on finding peaks in the \halpha\ emission, similar to what was done in \citet{WeiMonVer18}. We run the publicly available \textsc{astrodendro} code\footnote{\texttt{http://www.dendrograms.org/}}. Given an intensity map, this code splits an image into leafs, branches and trunks (the leafs being the star forming regions in our maps). We run  \textsc{astrodendro} with \texttt{min\_delta} = 6.0 $\times 10^{-19}$ erg/s/cm$^{-2}$ and \texttt{min\_npix} = 10. However, in order to minimize overlap between DIG and star forming regions we require for the dendrogram method that the DIG is at least 4 pixels away from the star forming regions. 

\begin{figure*}
 \includegraphics[width=2\columnwidth]{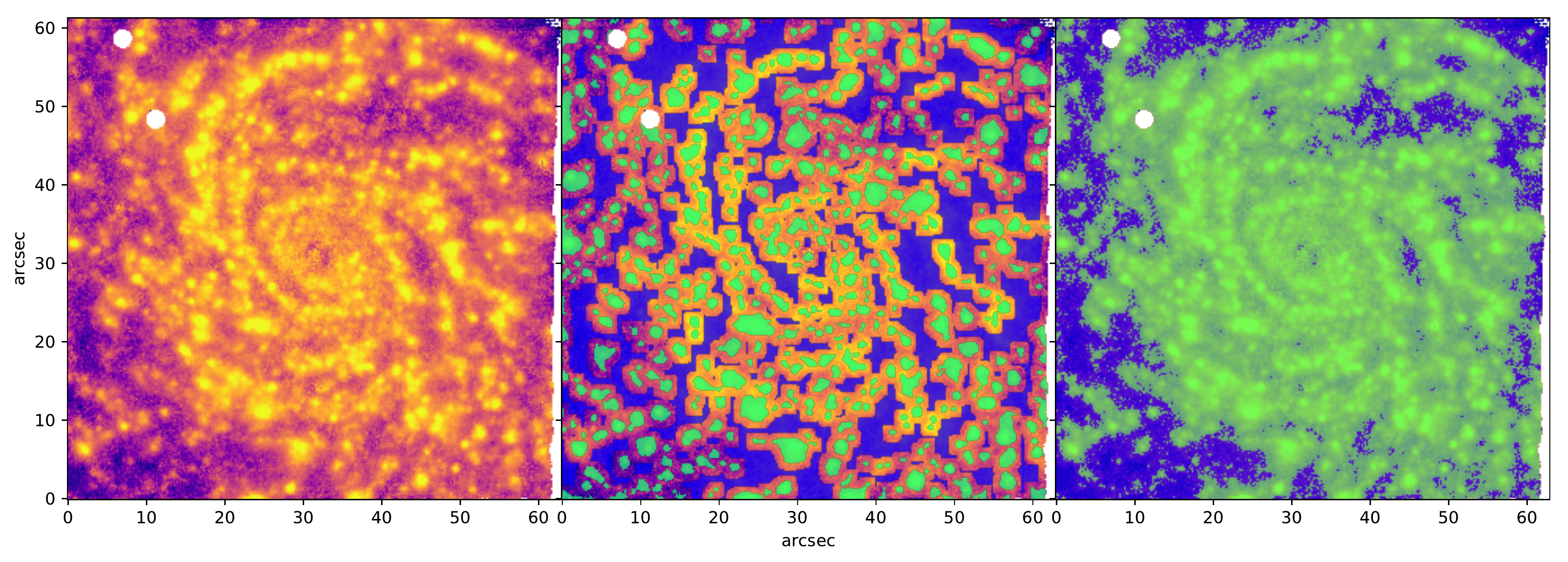}
 \caption{{ Identification of DIG and SF gas for NGC 4030. The left panel shows
   the \halpha map of NGC 4030. In the middle panel (dendrogram method) and
   right panel (Blanc method), the SF
   and DIG gas are identified as bright green and blue, while fully
   transparent areas are excluded.}}
 \label{fig:ngc4030_dig_id}
\end{figure*}

In Fig. \ref{fig:ngc4030_dig_id} we show for NGC 4030 which pixels are
identified as SF gas, DIG and neither for both methods. Similar Figures can be
found in Appendix F (online version only) for the five other galaxies in the kinematic subsample.

In Appendix F (online version only) we also compare the light and pixel fractions
  of DIG and SF gas. We find that the Blanc et al. method gives a relatively
  low fraction of pixel and light in DIG. The dendrogram method points at on
  average 30\% of the luminosity and 60\% of the pixels being in the form of DIG.

\subsection{DIG velocity subsample}\label{sec:subsample}
One of our goals is to derive rotation velocities for DIG and compare these with the velocities of SF gas. In order to carry this analysis out, we only look at galaxies for which the gas appears to be on circular orbits. For several galaxies, we see very complicated velocity structure close to their centres, for example the S-shapes seen in NGC 4941 and several other galaxies. In order to avoid very detailed modeling of the gas flows in these centers, we focus, for the {\it velocity} studies in this paper, on 6 galaxies without obvious kinematic twists and no large variation of systemic velocity of the gas with radius. The kinematic parameters of these 6 galaxies are summarized in Table \ref{tab:subsample}. Although we tried to avoid galaxies with bars, NGC 7162 might be weakly barred \citep{ButSheAth15}.

\subsection{Kinematic analysis}
In order to derive the kinematics we need to know the location of the
kinematic center, the kinematic position angle, and the flattening 
  ($1-b/a$, with
$a$ and $b$ the length of the major and minor axes) of the galaxy. Two of those quantities, the kinematic center and the flattening, are often not easy to determine. The position of the kinematic center is difficult to determine  because the motion of the gas is often irregular close to the center. The flattening is also difficult to infer from the kinematics, particularly for galaxy centres such as those that we are looking at.

We use \textsc{kinemetry} \citep{KraCapdeZ06} to fit radially the best fitting
ellipse with free flattening and kinematic position angle. For the kinematic center, we chose to adopt the photometric center of the galaxy, which we determine on the white light image of the MUSE cube.

For the kinematic position angle, we use the median value of the position
angle in the outermost 15\arcsec. The inclination is difficult to determine
from our data; the radial profiles derived with \textsc{kinemetry} show
significant scatter. We therefore use inclination values based on the
  axis ratios from the S4G photometry \citep{SalLauLai15}. These values are
  given in Table \ref{tab:subsample}. We estimate the inclination $i$ through
  $\cos(i) = q$, with $q = b/a$. This assumes an infinitely thin disk; assuming an intrinsic
  thickness of $q_0 = 0.2$ leads to an increase in inclination of 4 degrees
  for the most flattened galaxy.

We measure the rotation velocity of the diffuse and star forming gas in elliptical annuli defined by the position angle and flattening of the galaxy. Each annulus is 3\arcsec\ wide, as a compromise between radially blurring the rotation signal and having sufficient bins to infer the rotation velocity for each of the components. We then calculate the angle of each bin along the ellipse and determine the amplitude (and uncertainty on the amplitude) of the sinus that best fits the velocities along the ellipse using \textsc{lmfit} \citep{NewSteAll14}, taking into account the uncertainties on the bin velocities. We do this for both the DIG bins and the star forming bins. For the velocity dispersions, we calculate the maximum likelihood estimate of the  dispersion, without any angular dependence.

In order to compare with lower resolution data, we re-bin the 6 data
  continuum subtracted data cubes of the kinematic subsample to a sampling of
  0\farcs6, which we subsequently smooth to a resolution of 2\arcsec. Contrary
  to before, we do not distinguish between DIG and SF gas for
  these reduced-resolution data,  but instead measure the velocities and
  dispersions in a luminosity weighted way.

\section{Results}\label{sec:results}
\begin{figure*}
  \includegraphics[width=\textwidth]{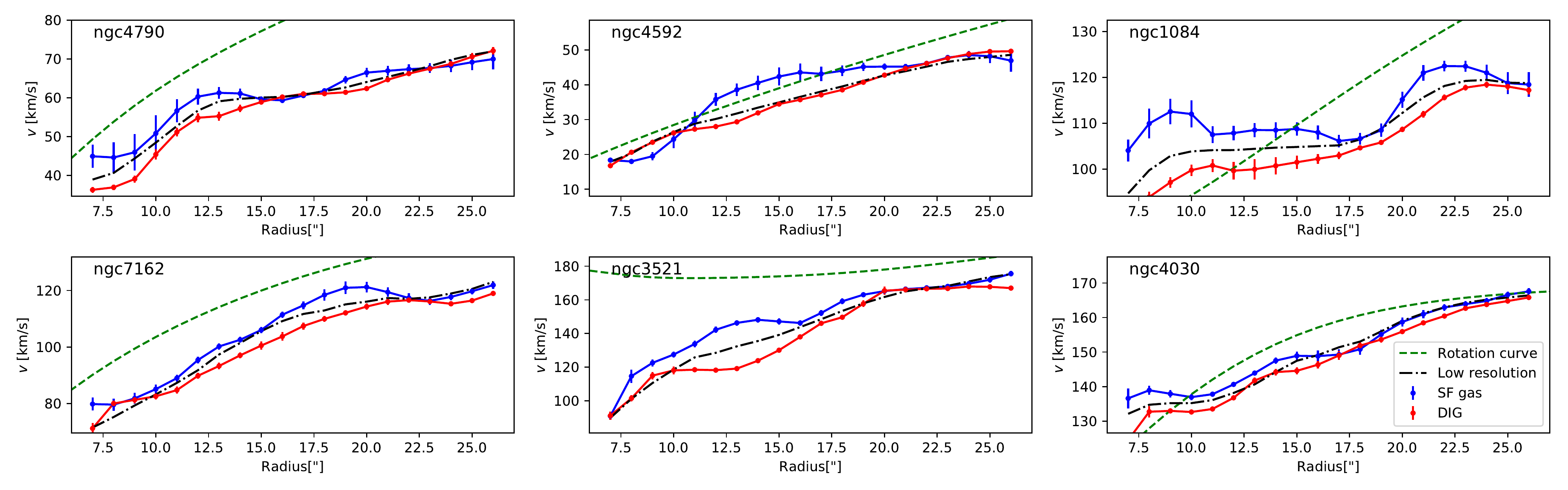}\\
  \vspace{0.5cm}
   \includegraphics[width=\textwidth]{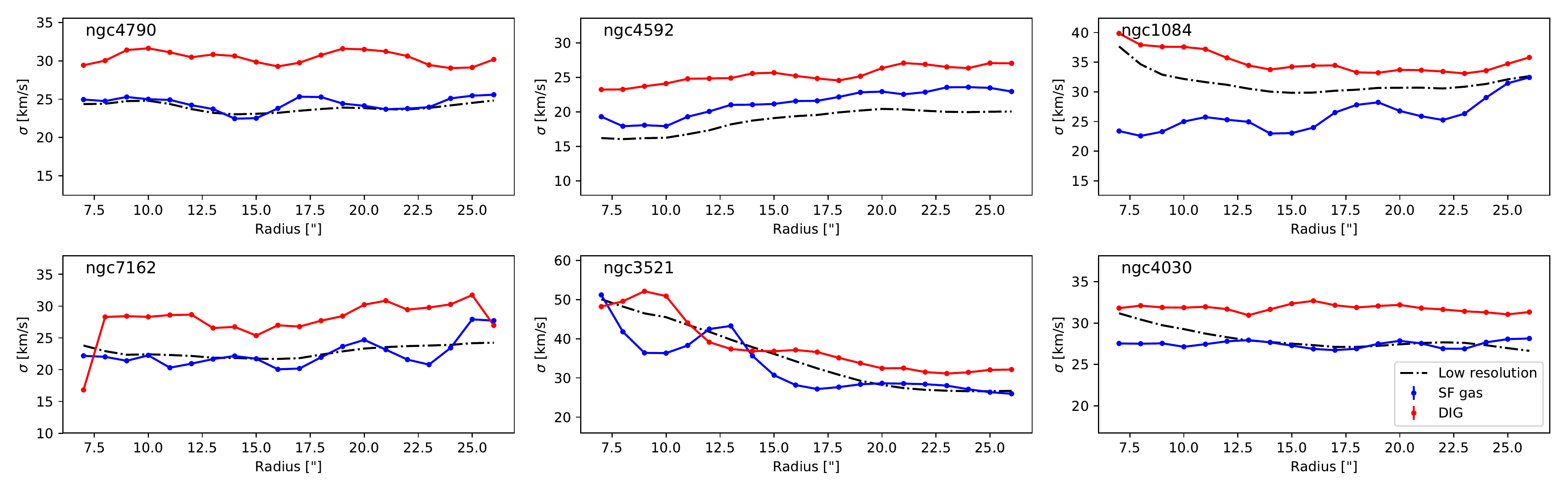}
 \caption{Rotation velocities (upper 6 panels) and velocity dispersion (lower
   6 panels) of star forming and diffuse gas in a subset of the 6 most
   regularly rotating galaxies in our sample. Star forming gas is shown in
   blue, diffuse ionized gas in red. The low-resolution luminosity weighted
     quantities are shown in black dashed-dotted lines.  Star forming and diffuse gas were separated by the dendrogram method. For completeness, we also indicate with green the circular velocity curve as derived from stellar kinematics (see Appendix \ref{sec:rotcurve}).}
 \label{fig:vel_dig_dendro_ext}
\end{figure*}

\begin{figure*}
  \includegraphics[width=\textwidth]{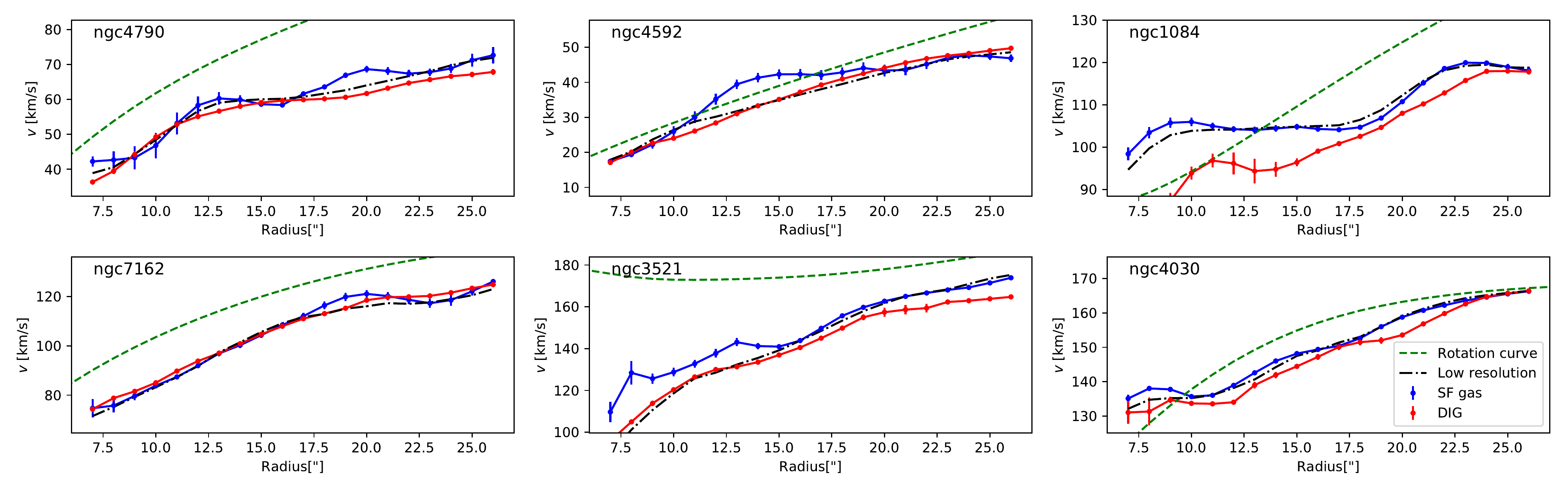}\\
  \vspace{0.5cm}
   \includegraphics[width=\textwidth]{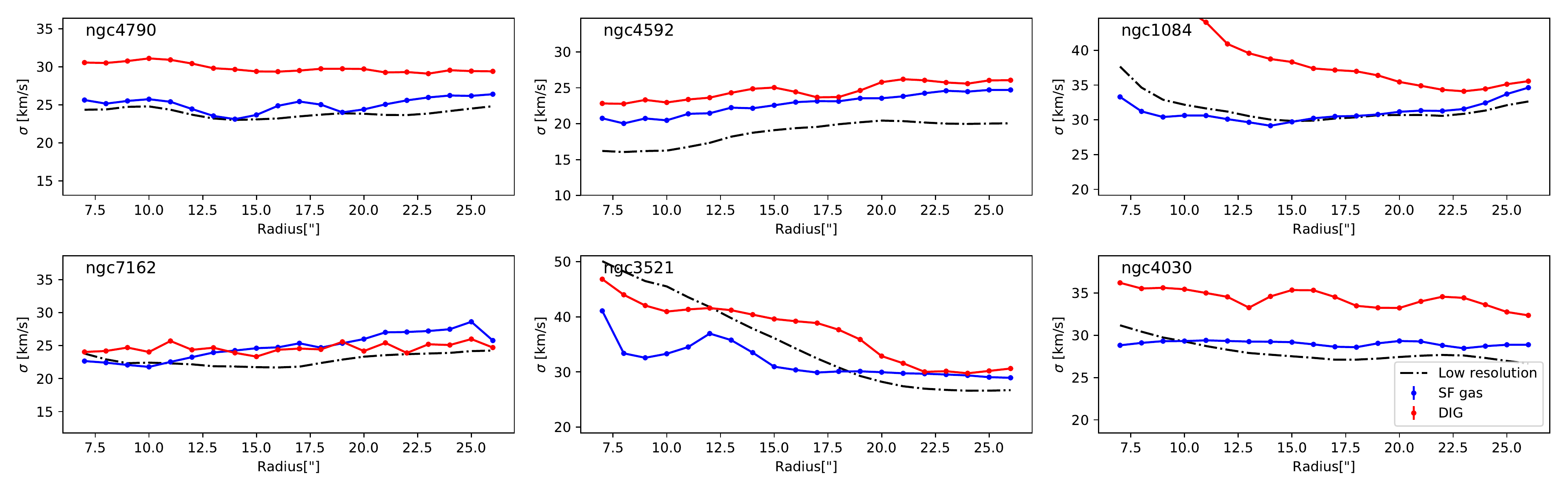}
 \caption{Rotation velocities (upper 6 panels) and velocity dispersion (lower 6 panels) of star forming (blue) and diffuse gas (red) in a subset of the 6 most regularly rotating galaxies in our sample. Same as Fig. \ref{fig:vel_dig_dendro_ext}, except that star forming and diffuse gas were separated by the [\ion{S}{2}]/\halpha surface brightness criterion.}
 \label{fig:vel_dig_f0_ext}
\end{figure*}

In Figs. \ref{fig:vel_dig_dendro_ext} and \ref{fig:vel_dig_f0_ext} we show the rotation velocities and dispersions of star forming and diffuse ionized gas. For both DIG identification methods, the DIG seems to be, on average, lagging behind the star forming gas. This is expected if we are indeed observing extraplanar gas.

The difference between the velocity dispersion of diffuse ionized gas and star
forming gas is more pronounced. (Note that the light-weighted
  low-resolution dispersion
  values can be lower than either measurement of the Voronoi binned gas as the
  convolution can spread out a single narrow peak over many pixels). We also
  note that the SF gas does not always agree with the rotation curves based on
stellar kinematics. Given the simplistic derivation of the latter, one does
not expect these curves to agree in detail; however, the SF gas rotating
slower in most cases might point at the necessity of an asymmetric drift
correction also for this component. The measured velocity dispersion, which is already corrected for instrumental broadening, consists of a thermal component and a gravitational component. As the thermal broadening is typically of the order of 10 km/s, most of the broadening must be due to gravitational broadening.

\begin{figure}
  \includegraphics[width=\columnwidth]{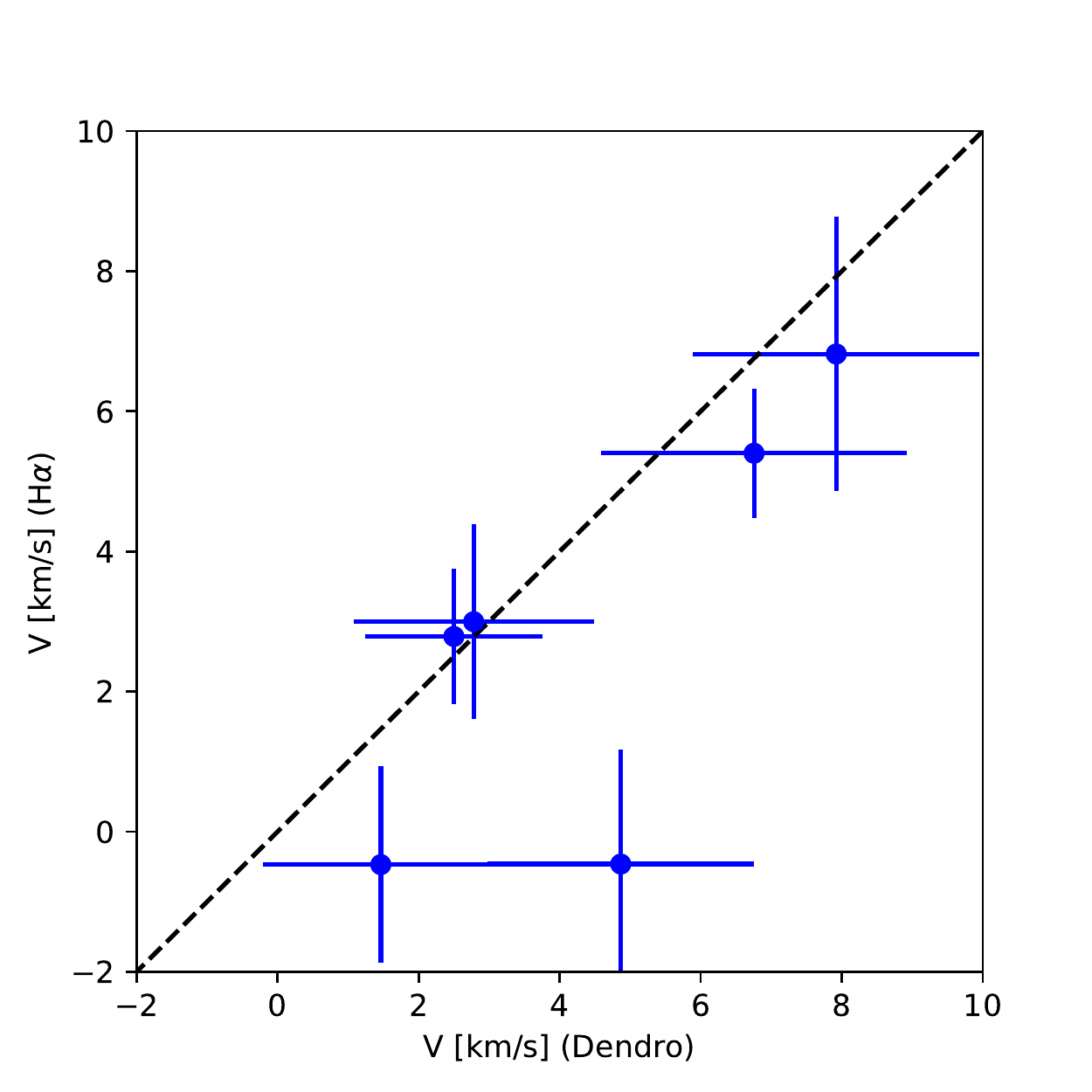}
 \caption{Median velocity difference between star forming and diffuse gas for the dendrogram method and the surface brightness method. For the dendrogram method all median velocity differences are higher than zero, meaning that the diffuse gas always lags with respect to the star forming gas.}
 \label{fig:vel_dig_v_v}
\end{figure}

\begin{figure}
  \includegraphics[width=\columnwidth]{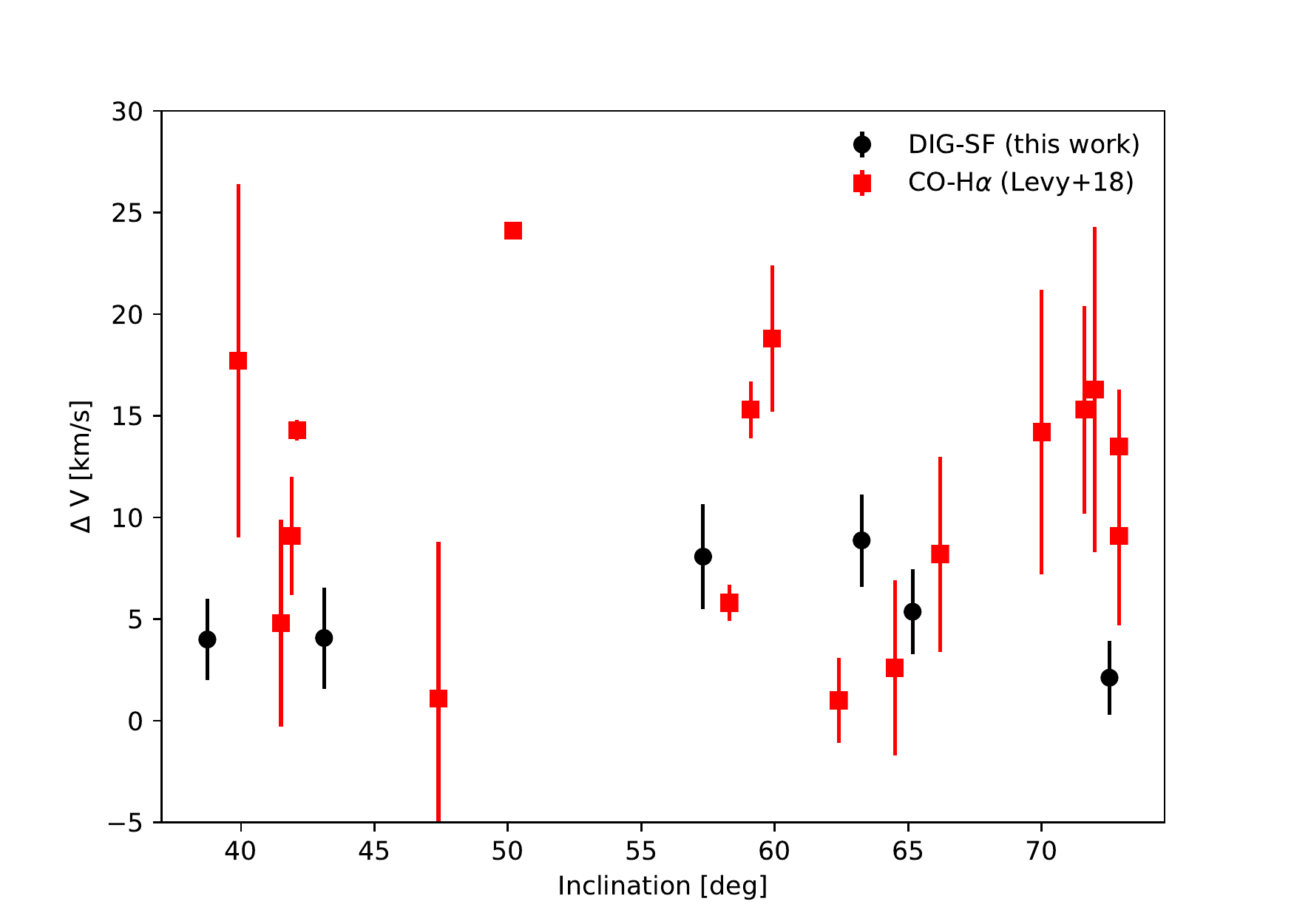}
 \caption{Median velocity differences for SF and DIG gas versus the
   inclination of the host galaxy (black). We also show the difference in CO
   velocity and \halpha\ velocity as determined by Levy et al. }
 \label{fig:vel_vs_levy}
\end{figure}

In order to quantify the difference in rotation velocity between the two
components, we follow \citet{LevBolTeu18} and take the median difference
between the rotational velocities of the DIG and star forming gas as well as
the median error on these differences. We note that taking a
uncertainty-weighted average gives similar values. In
Fig. \ref{fig:vel_dig_v_v} we show these median differences in rotational
velocity between the diffuse ionized gas and the star forming gas for both DIG
identification methods. For the dendrogram method, the velocity difference is
consistently higher than zero, meaning that star forming gas rotates faster
than DIG. For the surface brightness cut method, two galaxies (NGC 7162 and
NGC 4592) do not show any noticeable difference. In
Fig. \ref{fig:vel_vs_levy}, we compare the lags found with the dendrogram
method to the lags observed between \halpha\ and CO measurements by Levy et
al. Our difference measurements are on average 5 km/s lower than between
\halpha and CO. A Kolmogorov-Smirnov test suggests that the velocity
differences in this work and those of Levy et al. are not drawn from the same
sample, although with a relatively low significance level of $\alpha=0.05$. We
note that the average differences between the integrated low-resolution  measurement (black dash-dotted
curves in Fig. \ref{fig:vel_dig_dendro_ext}) and the star forming gas, which
are the two quantities most equivalent with the \halpha\ and CO measurements
of Levy et al., are even smaller. Although it is possible that for the
most-inclined galaxies the \halpha\ kinematics are affected by extinction and
therefore miss the fastest rotating gas near the midplane, we
see that (although low in number) also the less inclined galaxies in our
sample show on average lower values. 

Since the velocity fields are quite irregular, it is difficult to carry out the same analysis for the full sample. We therefore decided to only look at differences in dispersion between DIG and SF gas. To minimize the influence of bulges, bars and AGN, we only look at radii outside 14\arcsec. We exclude NGC 5643, IC 2560, NGC 4941, NGC 3081, NGC 4593 and NGC 3393, NGC 1566, NGC 1097 which are known AGN, and for which the BPT diagram (see Paper 2) shows that most of our gas observations are dominated by the AGN.

\begin{figure}
  \includegraphics[width=\columnwidth]{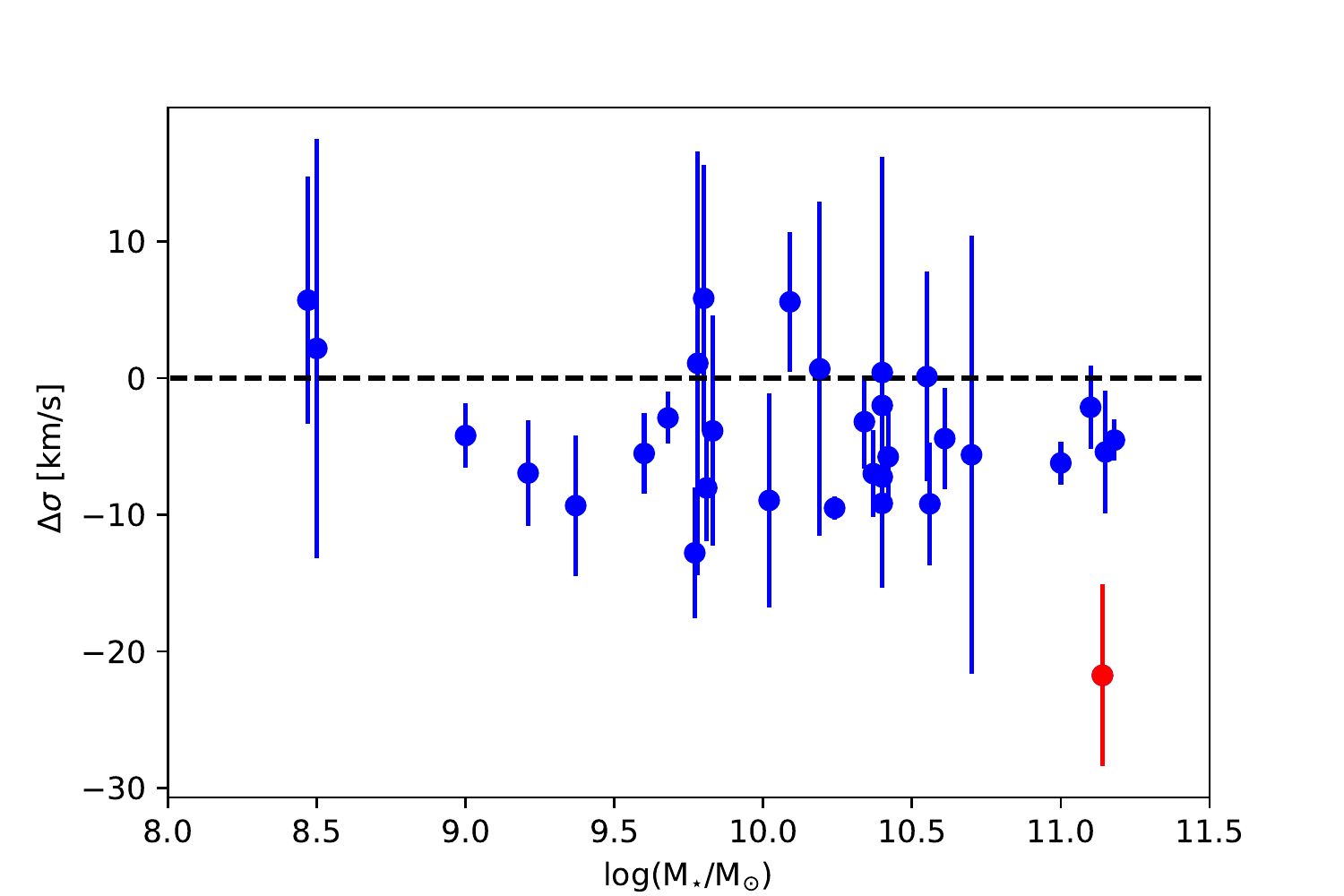}
  \includegraphics[width=\columnwidth]{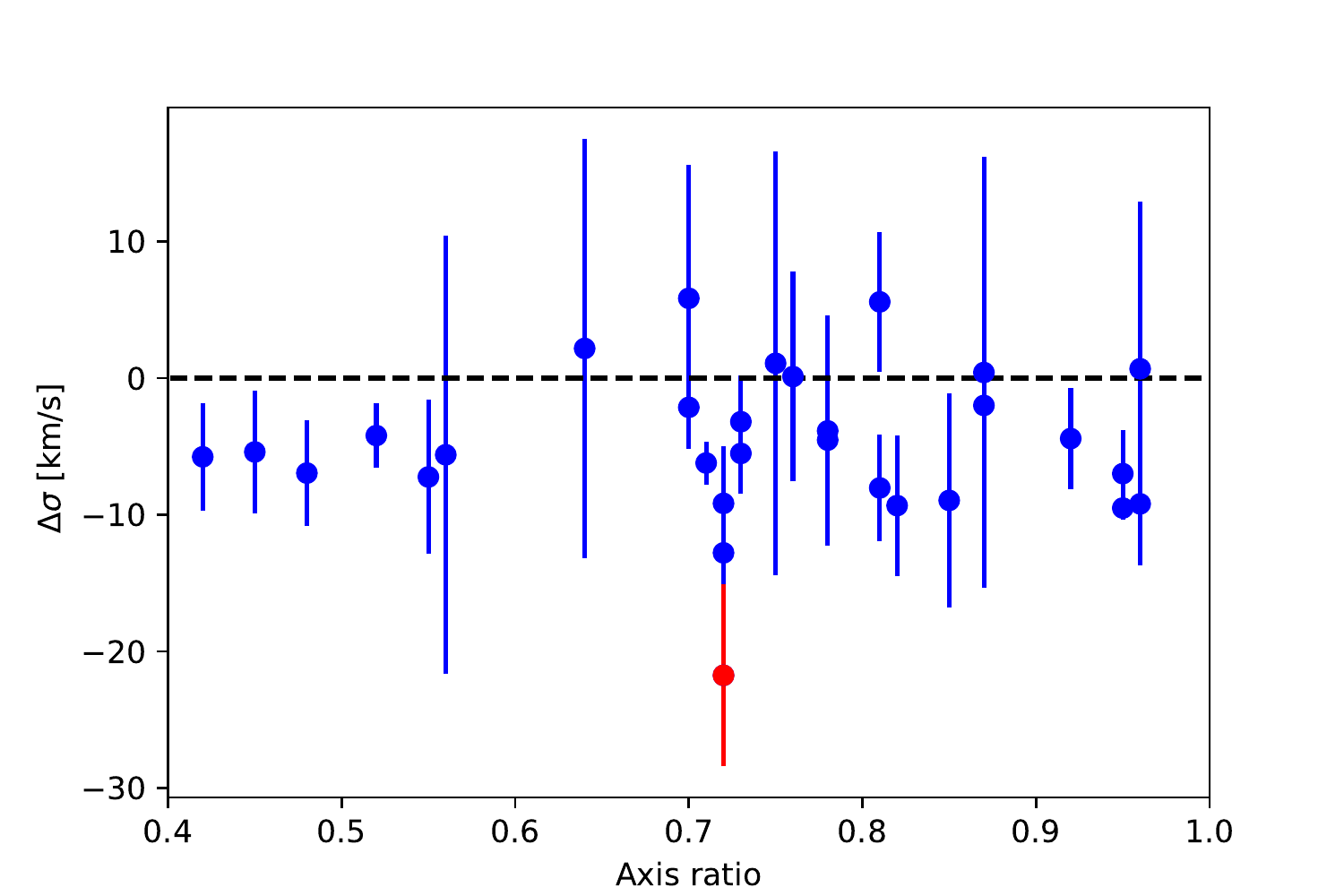}
 \caption{Median velocity dispersion differences for SF and DIG gas for the star forming sample as a function of stellar mass (upper panel) and axis ratio (lower panel). The red point is the merger remnant NGC 3256. A negative $\sigma$ means that the dispersion of the DIG is higher than that of the SF gas.}
 \label{fig:vel_dig_all_sigma_mass}
\end{figure}

In Fig. \ref{fig:vel_dig_all_sigma_mass} we show how the median difference in velocity dispersion depends on the stellar mass and the axis ratio. We note that on differences in velocity dispersion are consistent with being zero or lower than zero, meaning that the SF has a equal or lower dispersion than the DIG. The uncertainty weighted sample average is $\Delta\sigma = -6.9 \pm 2.7$ km/s.

If the gas we identified as DIG as indeed coming from extraplanar gas, one might expect a correlation between the mass of the galaxy and the dispersion of the dIG, since more massive galaxies have more massive disks (and higher surface mass density). From Fig. \ref{fig:vel_dig_all_sigma_mass} we note that there is also no evidence for a trend with stellar mass; once removing the point with the highest difference in $\sigma$, the merger remnant NGC 3256, the dependence on mass looks flat. 

We perform a similar check by looking at the differences in dispersion versus the axis ratio of the host galaxies. Assuming that all the galaxies are axially symmetric (which is not the case for NGC 3256), this is a proxy for the inclination of the host galaxy. Some DIG models \citep{MarFraCio10} allow for anisotropic velocity dispersions. We do not see any evidence from our data that the dispersions of the DIG depend on the inclination of the galaxy.

\section{Discussion}\label{sec:discussion}

\subsection{Possible influences on the measurement}
\subsubsection{Continuum subtraction}\label{sec:contsub}
We have shown in the previous sections how DIG can be seen to rotate slower than SF gas. We also showed that the strength of this rotational lag signal is dependent on the used method to identify DIG. Both methods however compare higher S/N data with lower S/N data. Although we do not believe that the Gaussian line fitting would lead to lower velocities for low S/N data, it is possible that during the fitting of the stellar continuum template mismatch alters the velocity of the \halpha\ line. As the stars generally rotate slightly slower and have a somewhat higher dispersion, underestimation of \halpha\ absorption in the stellar continuum could induce fake line emission at that location.

\begin{figure}
  \includegraphics[width=\columnwidth]{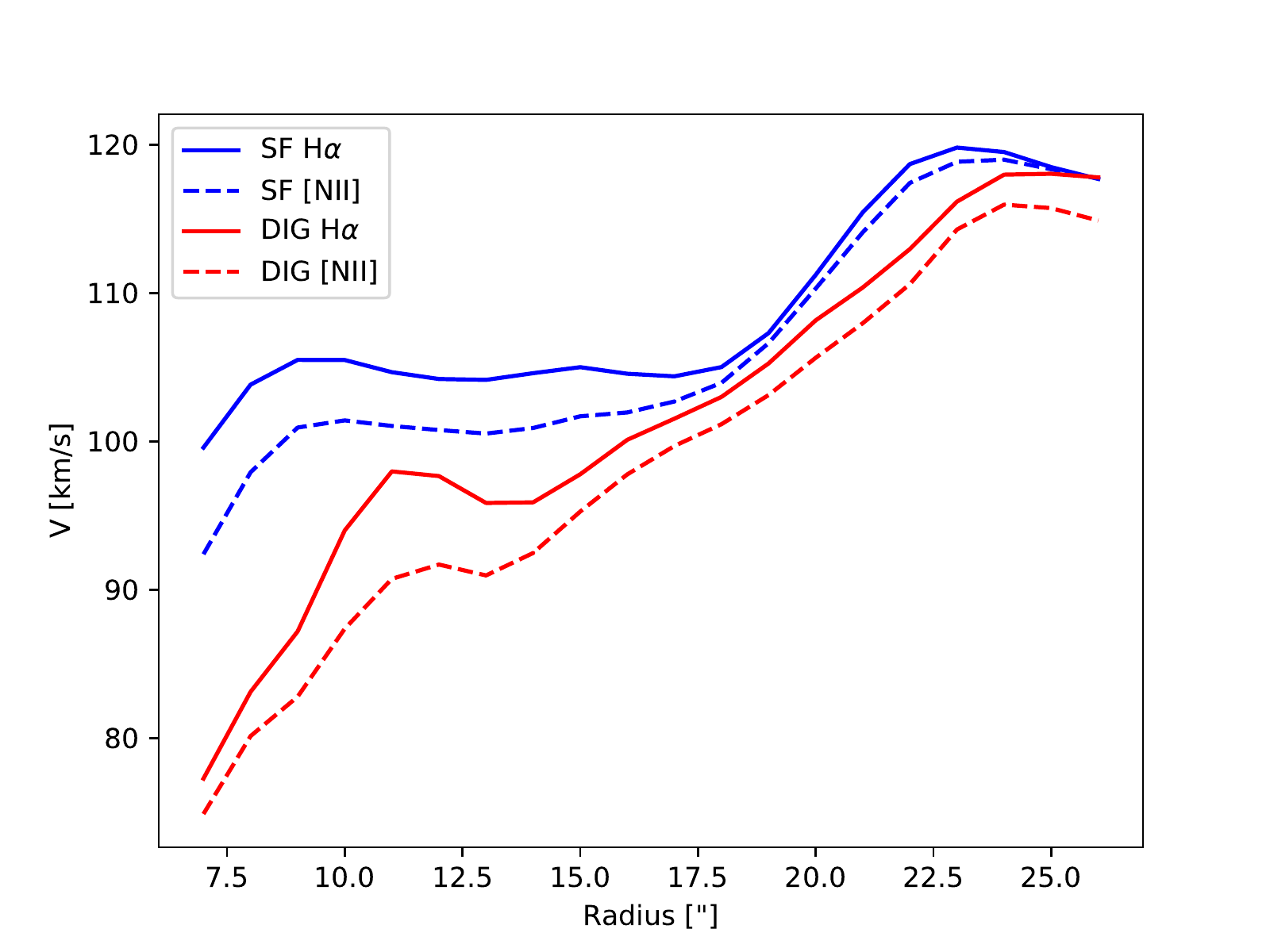}
  \caption{Comparison between the rotation velocities determined from the Balmer lines and the [\ion{N}{2}] lines for NGC 1084. Star forming gas is shown in blue, DIG in red. Velocities derived from  [\ion{N}{2}] lines are dashed. }
 \label{fig:comp_nii_halpha}
\end{figure}

To show that this is not the case, we repeat the analysis for the 6 galaxies with DIG identification by \halpha\ surface brightness, but instead of fitting the velocity of the Balmer lines, we fit the velocity of the two [\ion{N}{2}] lines at 6548 and 6583 \AA. These two lines are much less sensitive to continuum subtraction problems, as they are on opposite sides of \halpha, while they are still bright enough to accurately determine velocities. In Fig. \ref{fig:comp_nii_halpha} we show the rotation velocities of star forming and diffuse gas as determined from Balmer and the [\ion{N}{2}] lines for NGC 1084. We have checked the other 5 galaxies and find the same behaviour there. The [\ion{N}{2}] kinematics are different for SF gas and DIG, with the SF gas rotating faster than the DIG. This confirms our suspicion that the measurement is not affected by the continuum subtraction. For the gas identified as SF gas, the rotational velocities are lower for the [\ion{N}{2}] lines than for the Balmer lines. We think that this is because the relative contribution of the DIG to the observed flux is higher for  [\ion{N}{2}], as the ratio of [\ion{N}{2}]/\halpha\ is known to enhanced for DIG \citep[e.g.][]{HafReyTuf99,HooWal03,MadReyHaf06}.

\subsubsection{Extinction correction}
We correct the \halpha\ fluxes for dust attenuation using the Balmer decrement, assuming an intrinsic ratio of 2.86 and a \citet{Fit99} dust law, before identifying the DIG. It is possible that the light of a heavily attenuated HII region mixed with the light of the DIG with different kinematics, will lead to an observed region with SF-like brightness but DIG-like kinematics. We have therefore checked if the results are significantly different if we would not correct for extinction. For this, we have re-calculated the $f_0$ values in the same way as in Sec. \ref{sec:id_dig}, but without extinction correction, and re-measured the kinematics for the new divisions between DIG an SF gas. Although we see some small quantitative differences, we do not see any qualitative differences in the results.

\subsection{Most of the gas we identify as DIG is close to the midplane}
In this Section we explore the possible vertical distribution of the DIG in the 6 galaxies. We argue that both from a velocity point of view, as well as from a dispersion point of view, our data suggest a majority of the DIG emission that we are seeing is close to the midplane. For this, we assume that the Jeans equation holds, and that the gas and the potential follow a vertically exponential distribution. 

1) For the two DIG identification methods used in this paper, the velocity lags are modest. Using the kinematic separation technique, we find velocity differences that are slightly higher than the other two methods. We use Eqn. \ref{form:vlag} (see Appendix) as a measure to identify the scale height of the DIG normalized by the scale height of the potential $h_z$. We assume that the velocity of the star forming gas is equal to the rotation velocity in the midplane, and that the scale height of the potential is close to the scale height of the stars. A 10\% lag in velocity would then lead to a scale height that is 20\% of that of the stars (note that for a self consistent potential this leads to the stars rotating at $\frac{2}{3} V_0$, with $V_0$ the circular velocity in the midplane). Using a measurement for the lag in the velocity may actually in some instances be a more solid indicator of the spatial distribution of the gas than the velocity dispersion. For the latter, it is not always known what the contribution of non-gravitational motions is, whereas for a velocity measurement such components would most likely average out. However, we do acknowledge that there is still considerable uncertainty in interpreting these lags, as there is no guarantee that the gas is on circular orbits.
We convert the velocity lags to scale heights for the velocity lags found by separating DIG and SF gas with the dendogram method. With the exception of two peaks in NGC 4592 and NGC 3521, the inferred scale heights are very moderate and always much lower than the scale height of the stars. 

2) The dispersions can provide an independent measurement of the vertical scale height (Eqn. \ref{form:siglag}). Using the median dispersion ratios between DIG and SF gas from the dendogram method, we find that dispersions of the DIG are 16\% (NGC 4030) to 35\% (NGC 1084) higher than that of the SF gas. At face value, this would lead to scale hights of the DIG that are similar fractions higher, and therefore only marginally higher than the scale height of the SF gas. However, it is not known what the contribution of non-gravitational motions is to the velocity dispersion of the gas, and therefore this fraction can be much higher.
Using an estimate for the stellar scale height and stellar surface mass density in Eqn. \ref{form:siglag}, we can obtain an upper limit on the scale height. From the pPXF fits to the stellar continuum, we can obtain a crude estimate of the stellar surface mass density. The details of this procedure are outlined in Paper 2. The templates we use are based on a Kroupa IMF. We note that using a Salpeter IMF would lead to a higher surface mass density and therefore a lower scale height of the DIG.

\begin{figure*}
  \includegraphics[width=\textwidth]{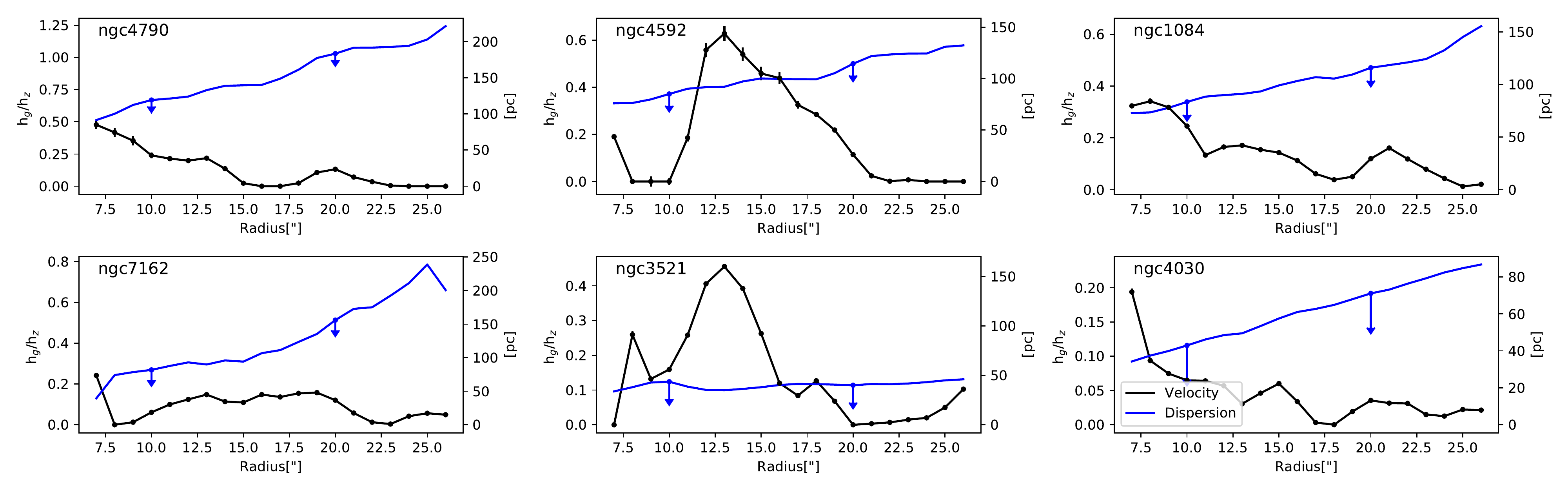}
  \caption{Radially varying scale heights of DIG as inferred from velocity
    differences (black) and dispersions {(blue) with respect to the scale
      height of the stellar disk. The scale height in pc is
      given on the right axis}. As the non-gravitational component of the dispersions is not known, the dispersion measurements are approximate upper limits. }\label{fig:scaleheight}
\end{figure*}

In order to infer the vertical distribution of the stars, we make use of the
relation between scale length and scale height of disks given by Eqn. 1 in
\citet{BerVerWes10}. The stellar scale heights for all 6 galaxies vary
between 178 pc (for NGC 4790) and 371 pc (for NGC 4030). Solving
Eqn. \ref{form:siglag} with these values, and using the radial dispersion
profiles for the DIG as identified from the dendrogram method, we find upper
limits to the scale heights of the DIG. These are shown in
Fig. \ref{fig:scaleheight}. We see that the upper limits increase with
radius. This is likely a consequence of using the stellar surface mass density
as the dominant part of the potential. It is expected that the potential at
larger radii will be more dominated by dark matter. As the scale heights for
the velocity lags are based on a Mestel disk with a constant rotation curve,
the scale heights based on velocity do not show radially rising profiles. Most
important is however that under the assumption of an exponential vertical
distribution almost all the upper limits on the DIG scale heights are
consistent with the DIG originating from a layer that has a lower scale
  height than the stars.

We note that there is an alternative way of identifying DIG, which consists of
performing double Gaussian component fits to emission lines. In Appendix
\ref{sec:twocomp}, we explore this method and see that our conclusions of most
of the DIG has a scale height lower than that of the stars is robust. 

\section{Summary}
In this paper we have presented the methodology for the kinematic extraction and the presentation of the kinematics maps of the first 41 galaxies of the MAD Survey with VLT/MUSE. We have outlined the methodology for the measurement of the gas kinematics, which we will use in future papers. 

We use two methods, one based on the ratio of [\ion{S}{2}]/halpha \citep{BlaHeiGeb09}, the other on finding peaks in the \halpha\ maps, to identify diffuse ionized gas. For a subsample of 6 regularly rotating galaxies it is possible to measure differences in rotational velocity between ionized and star forming gas. We find median velocity lags of 0-10 km/s. These measurements are dependent on the method used to identify DIG.

Using the dendrogram method we measure differences in ionized gas velocity dispersions for DIG and SF gas for the rest of the sample after removing galaxies affected by an AGN. Although we caution against possible biases in the velocity dispersion due to the limited spectral resolution of our data, we do find a consistenly higher velocity dispersion for DIG than for star forming gas. We do not find any dependence for this value on mass or inclination.

Interpreting the measured velocity lags and dispersions differences with Jeans
models, we show that the observed diffuse gas in our data is
consistently located close to the midplane of the galaxy,  with scale
  heights that are below those of the stars. This limits the models for the origin of DIG gas to processes such as expanding shells, leaky HII regions and possibly ionizing radiation from older stars.

\section*{Acknowledgements}
Based on observations made with ESO Telescopes at the La Silla Paranal Observatory under programme IDs 60.A-9100(C), 095.B-0532(A), 096.B-0309(A), 097.B-0165(A), 098.B-0551(A), 099.B-0242(B), 100.B-0116(A). We thank the ESO staff for their assistance during the observations. JB acknowledges support by FCT/MCTES through national funds by grant UID/FIS/04434/2019 and through Investigador FCT Contract No. IF/01654/2014/CP1215/CT0003. MdB thanks Kyriakos Flouris for sharing his reduction script and Sebastian Kamann for comments on the draft. This research made use of Astropy, a community-developed core Python package for Astronomy (Astropy Collaboration, 2013). This research made use of astrodendro, a Python package to compute dendrograms of Astronomical data (\texttt{http://www.dendrograms.org/}). This research has made use of the NASA/IPAC Extragalactic Database (NED) which is operated by the Jet Propulsion Laboratory, California Institute of Technology, under contract with the National Aeronautics and Space Administration.

%%%%%%%%%%%%%%%%%%%%%%%%%%%%%%%%%%%%%%%%%%%%%%%%%%



\bibliographystyle{mnras}
\bibliography{paper} % if your bibtex file is called example.bib

%\begin{thebibliography}{99}
%\end{thebibliography}

%%%%%%%%%%%%%%%%%%%%%%%%%%%%%%%%%%%%%%%%%%%%%%%%%%

%%%%%%%%%%%%%%%%% APPENDICES %%%%%%%%%%%%%%%%%%%%%

\appendix

\section{Simulations to check the recovery of the velocity dispersion}\label{apx:sim_gas}
For our analysis, we measure the position and the width of the Balmer lines and the [\ion{N}{2}] and [\ion{S}{2}] lines. However, most lines have widths that are well below the width of the line spread function of MUSE. This is generally not a problem to estimate the centroid of the line, however, it becomes increasingly difficult to estimate the dispersion correctly for low values of the dispersion. In order to assess how reliable the dispersion measurements are, we perform a small suit of simulations.

We focus exclusively on the Balmer lines, since we do not use the dispersion of the other lines. We generate \halpha\ and H$_{\beta}$ in a ratio of 3:1 on an spectrum that is oversampled 10 times. We assume idealized conditions to test the kinematics and therefore do not simulate the stellar continuum in the mock observations.

For each velocity dispersion we simulate 1000 spectra. We assume the emission
line spectrum is redshifted by a velocity of 1600 km/s, which is typical of
galaxies in the sample, and we additionally add  a random velocity drawn from
a Gaussian distribution with a width of 10 km/s. The width of the Gaussian is
obtained by taking the squared sum of the line spread function and the
dispersion. 
It is known that the LSF varies spatially. In order to capture the
  spatial variation in our simulations, we use a reduced cube of arc lines. As
  our data are the average of 4 different rotator angles, we use the same
  rotation angles to generate 4 arc cubes, which we than stack with small
  random Gaussian offsets ($\sigma=$1\arcsec). We find that the spatial
  variation of the FWHM of the LSF at 6595 \AA\ is 0.09 \AA. At the wavelength
  of \halpha this corresponds to an uncertainty of $\sigma$=2 km/s. This variation is
  higher than the one measured by \citet{GueKraEpi17}, likely because they
  average many more observations. We include this
  uncertainty to our simulations by adding this as a Gaussian random variable
  to the LSF.

After generating the two emission lines, we rebin the spectrum and
add Gaussian noise in such a way that the signal-to-noise is the same as what
we would measure from the Voronoi binning of the cube on the \halpha\ line.

We then fit the lines in the same way as we fit the data, except that we skip the continuum subtraction part. These simulations are thus very much idealized and present an upper limit for how well we understand the measured velocities and dispersions.

\begin{figure*}
  \begin{center}
\includegraphics[width=0.47\textwidth]{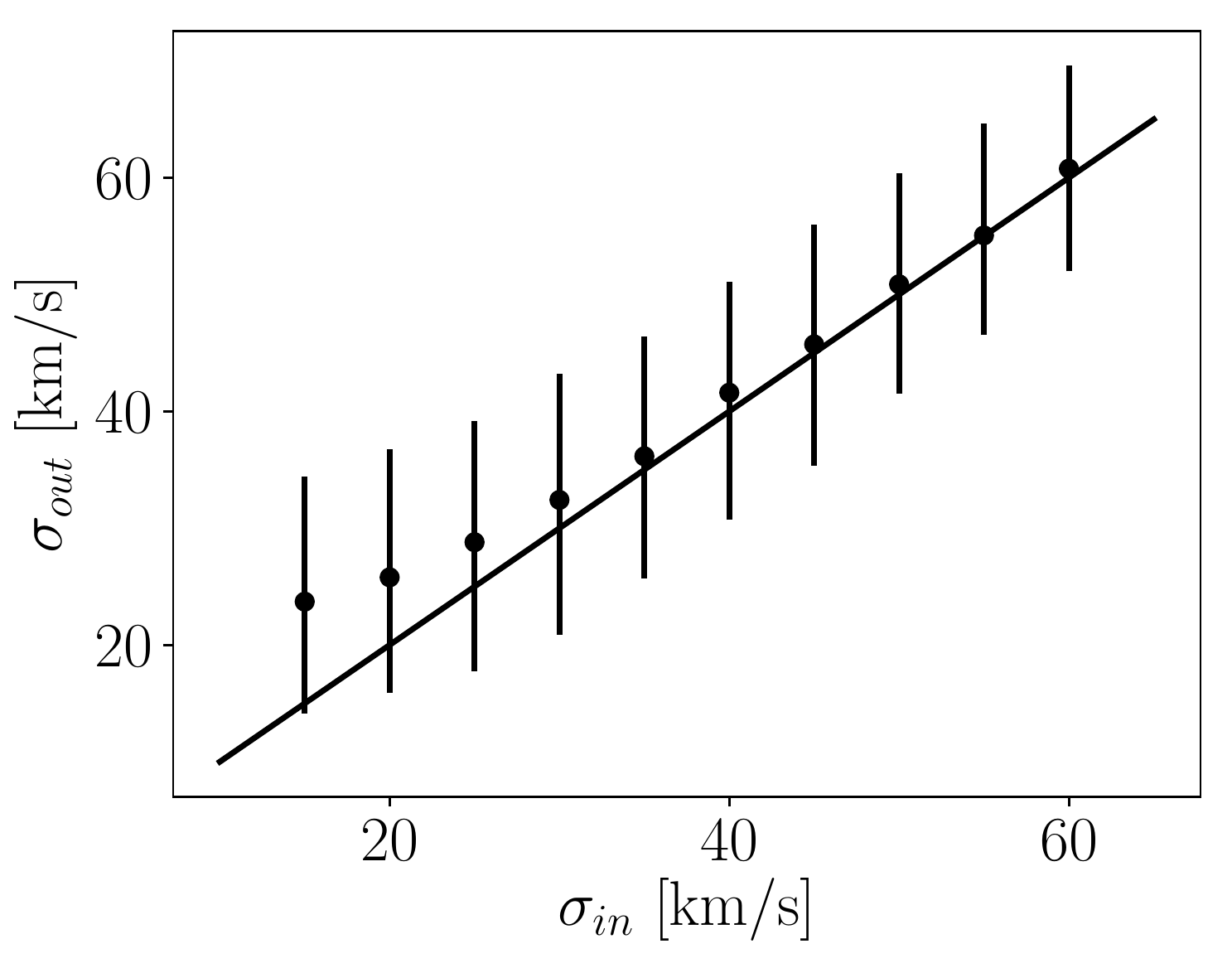}
\includegraphics[width=0.47\textwidth]{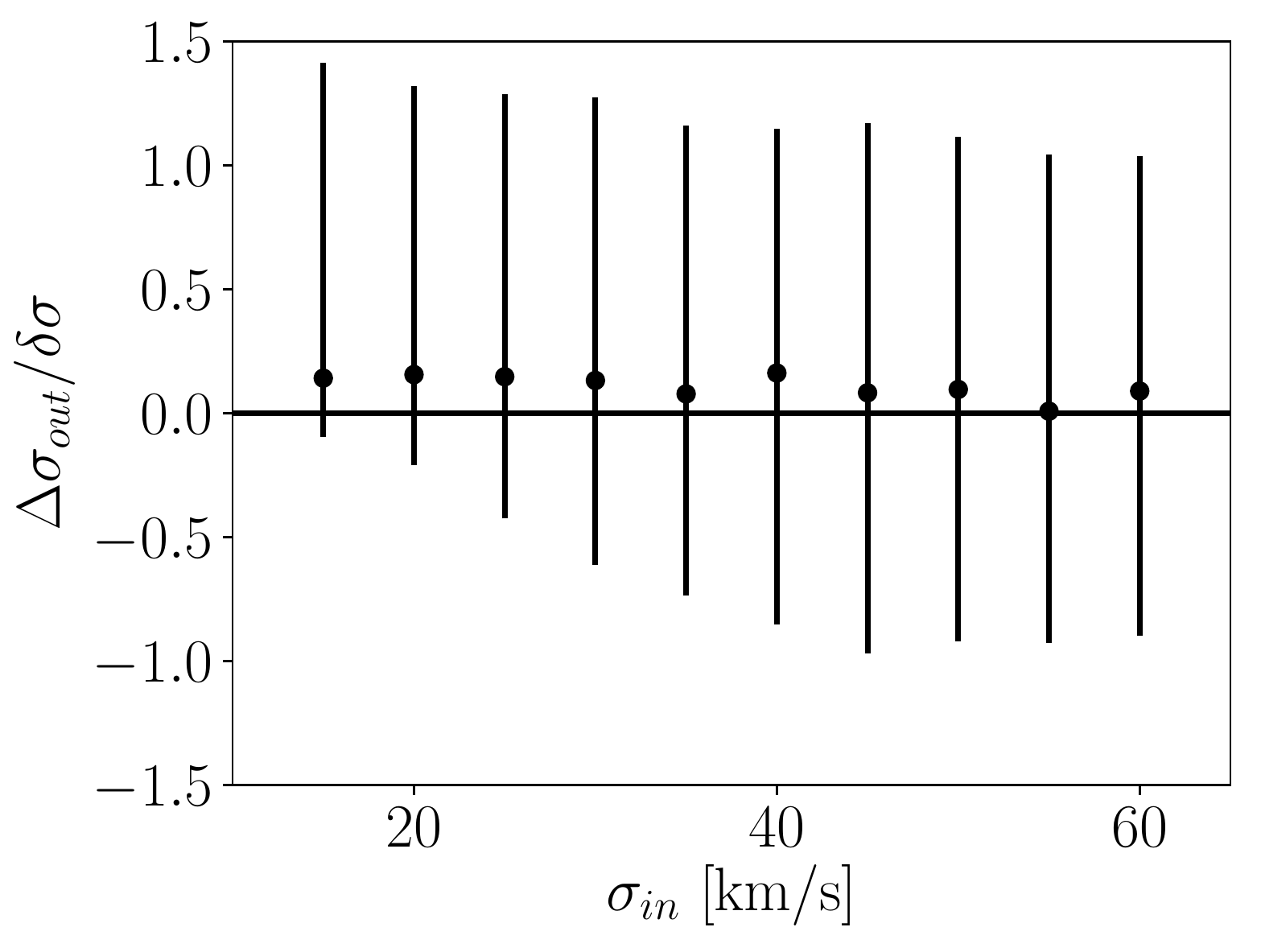}
\caption {Results from simulations for the recovery of the velocity
  dispersion. Left panel shows the recovered dispersion as a function of the
  input dispersion. Dispersions are unbiased until about 35 km/s, but recovery
  is biased for narrower lines toward higher dispersions. In the right panel
  we show the median bias in sigma ($\sigma_{out} - \sigma_{in}$) and the 68\%
  confidence interval normalized by the median error bar from the fitting
  procedure as a function of the input dispersion. As expected, the size of
  the error bars is close to 1 for high dispersion objects. Low
    dispersion objects deviate slightly because of the uncertainty in the LSF. }\label{fig:sim_gas_sigma_vs_sigma}
\end{center}
\end{figure*} 

Fig. \ref{fig:sim_gas_sigma_vs_sigma} shows the measured velocity dispersion
versus the input velocity dispersion and shows that we can recover unbiased
dispersions down to about 35 km/s. The width of the error bars shows the 16th
and 84th percentiles of the recovered dispersions. In the right panel we show
the bias in the recovery of the dispersion as a function of the input
dispersion. The error bars in the Figure show the  68\% confidence range and
are normalized by the median uncertainty as determined from the fit to the
simulated data at that input dispersion. The recovered uncertainties are
  close to 1 for high velocity dispersions, but become higher toward
    lower dispersions. This is consistent with the added uncertainty in the
    LSF.

\section{Jeans models for extraplanar gas}\label{apx:jeans}
 We assume that the density distribution of the galaxy is dominated by the stellar mass in the plane of the galaxy, and that the vertical distribution has an exponential profile. We can than write the physical density at radius $R$ and elavation above the disk $z$ as $\rho(R,z) = \frac{\Sigma(R)}{2 h_z} \exp\left(-z/h_z\right)$, where $h_z$ is the scale height of the stars, and $\Sigma(R)$ the surface mass density.

Under the assumption of a flat axially symmetric potential, we can write the Poisson equation close to the midplane as \citep[e.g. Eq. 2.74 from ][]{BinTre08}:
\begin{eqnarray}
\frac{\partial^2\Phi(R,z)}{\partial z^2} &=& 4\pi G \rho(R,z). 
\end{eqnarray}
From this, we derive an expression for the vertical component of the force:
\begin{eqnarray}
\frac{\partial\Phi(R,z)}{\partial z} &=& 2\pi G \Sigma(R) \left(1 - \exp(-z/h_z)\right).
\end{eqnarray}
Note that the integration constant here is zero, so that the force is zero exactly in the midplane. We combine this with the Jeans equation in the z-direction for an axisymmetric system in which the velocity ellipsoid is aligned to the symmetry axes \citep[e.g.][]{Cap08},
\begin{eqnarray}
\nu\frac{\partial \Phi(R,z)}{\partial z} &=& -\frac{\partial \nu\overline{v_z^2}}{\partial z},\label{eqn:jeans}
\end{eqnarray}
with $\nu$ is the (vertical) distribution function of the tracer population, which we assume to have the form $\nu(z) = \nu_0\exp(-z/h_g)$. This equation allows us to find an expression for $\nu\overline{v_z^2}$:
 
\begin{eqnarray}
\nu\overline{v_z^2} = \nu_0 2\pi G \Sigma(R) h_g \left[ e^{-z/h_g} - \frac{e^{-z(\frac{1}{h_g} + \frac{1}{h_z})}}{1+\frac{h_g}{h_z}} \right].
\end{eqnarray}
So that the expected value for the squared velocity dispersion becomes:
\begin{eqnarray}
\sigma_z^2  = 2 \pi G \Sigma(R) h_g \left[ 1 - \frac{1}{(1+\frac{h_g}{h_z})^2}    \right].\label{form:siglag}
\end{eqnarray}
Note that by taking $h_g = h_z$ this result simplifies to the well-known solution for hydrostatic equilibrium  for an exponential scale height \citep[e.g. Eqn 25 in ][]{van88}.

Disk galaxies are often modeled as a Mestel disk, as this potential
  reproduces the flat rotation curves observed in galaxies \citep[e.g.][]{BinTre08}. 
Assuming a Mestel disk, \citet{LevBolTeu18} derive the expected rotation velocity of gas at a altitude $z$ above the midplane from the disk:
\begin{eqnarray}
\frac{V(z)}{V_0} = \exp\left(-\frac{|z|}{2 h_z} \right),
\end{eqnarray}
with $V_0$ the circular velocity in the midplane. It is straightforward to show that if a gas disk has an exponentially declining vertical distribution and is sufficiently geometrically thin the observed average velocity of such a distribution will be:
\begin{eqnarray}
\overline{V} = V_0 \frac{1}{1 + \frac{h_g}{2 h_z}}. \label{form:vlag}
\end{eqnarray}

\section{Two-component fits}\label{sec:twocomp}
For the kinematic subsample for which we determine the rotation velocity of the DIG (Sec \ref{sec:subsample}), we also fit double Gaussian component line profiles to the continuum subtracted spectra. To do this, we first tesselate these data to a high S/N of 50 at the location of \halpha. We assume that each line is composed of a narrow and a broad component, and that in this case the kinematics of each component of the \halpha line and lines from [NII] and [SII] are the same: for each voronoi bin the narrow components of all emission lines have the same velocity $V_0$ and velocity dispersion $\sigma_0$ and similar for the broad components ($V_1$, $\sigma_1$). The one-component fits were carried out with a damped $\chi^2$ fitting method. For the two-component fits we chose to use a MCMC sampler \citep[emcee,][]{ForHogLan13}, as such a sampler is more likely to determine the global minimum.

\begin{figure*} 
  \includegraphics[width=0.45\textwidth]{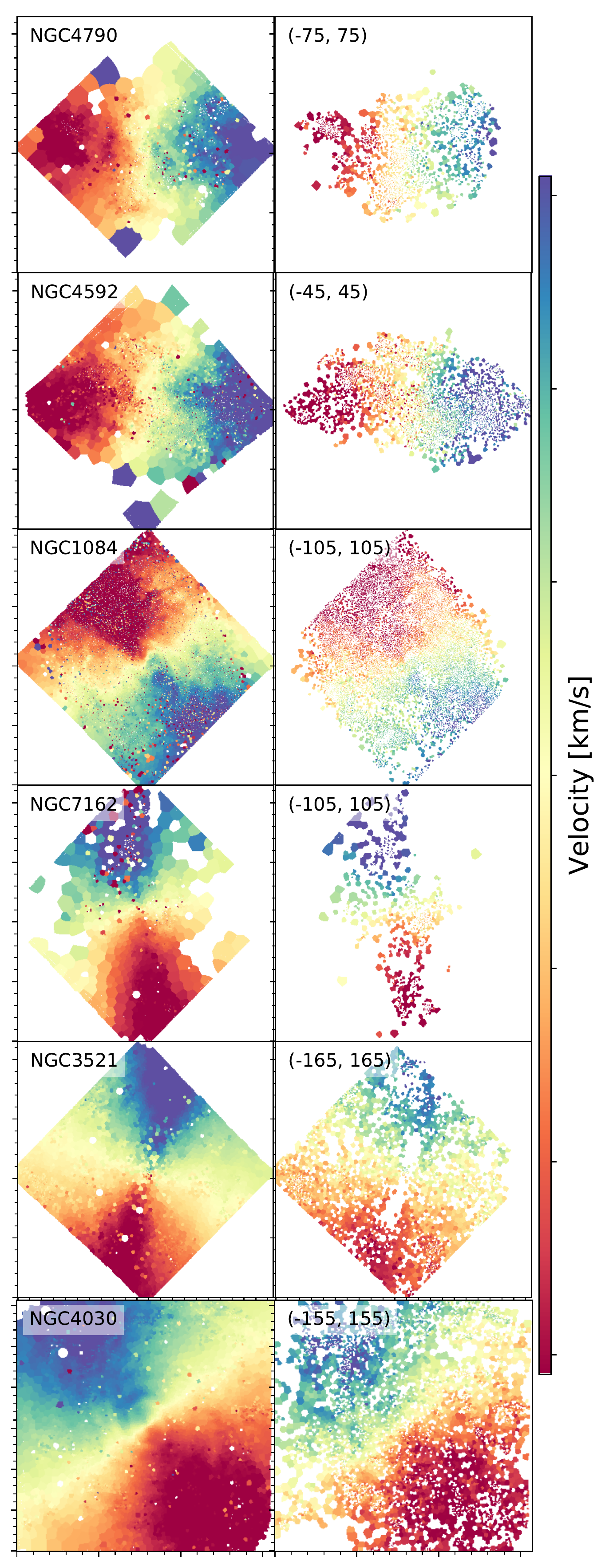}
  \includegraphics[width=0.45\textwidth]{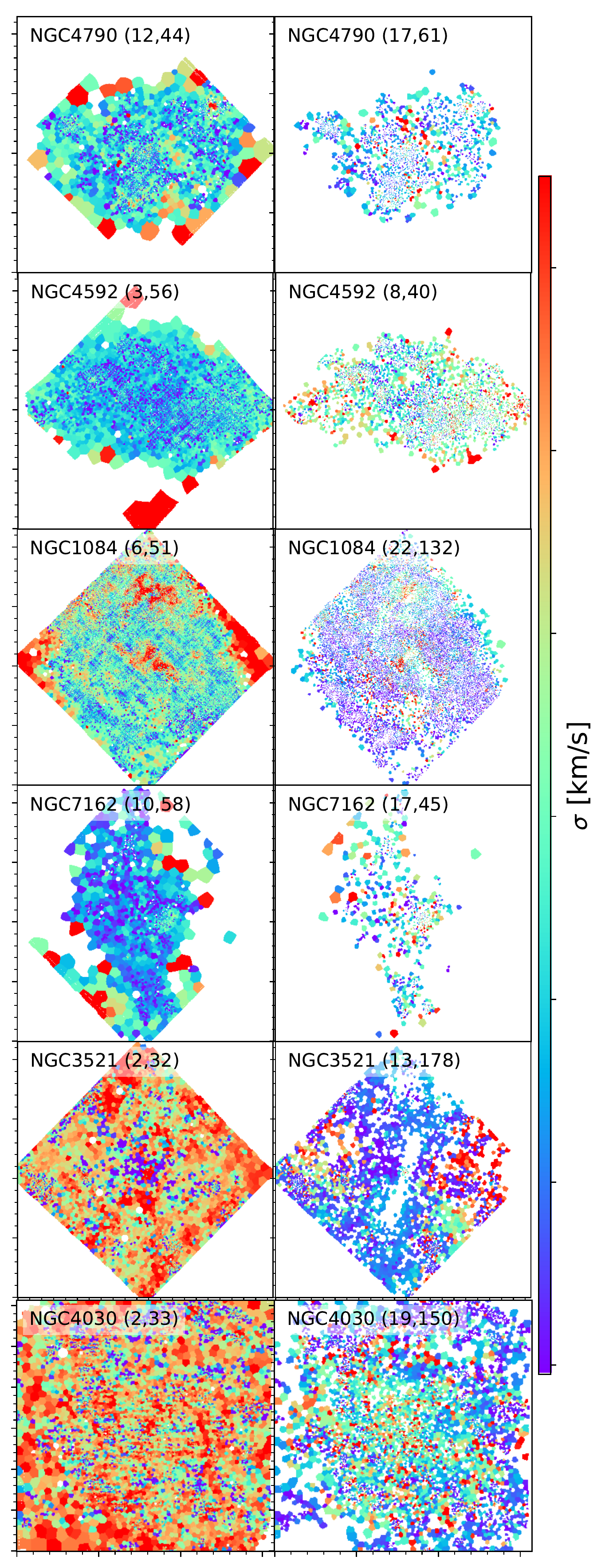}
  \caption{Maps of kinematics based on two-component fitting. Left panels show the velocity of the narrow component, right of the broad component. The range in velocity is the same for the narrow and broad component, and is given in the panels of the broad component. The dispersion is scaled per component between the 5\% and 95\% percentile, and are scaled linearly.}\label{fig:2comp}
\end{figure*}

We acknowledge that such a decomposition is difficult at the spectral resolution of MUSE. We have removed the fits for which we did not see any convergence in the velocity dispersion of the second component, by removing all fits with $\delta\sigma > 60$ km/s. Although these decompositions are uncertain (not only are velocities and dispersions between different components degenerate, but the fits are also sensitive to small residuals from the sky subtraction), they do provide an interpretation of the data that is not necessarily less valid than a single component fit. The kinematics of these decompositions of the 6 galaxies are shown in Fig. \ref{fig:2comp}.

It is interesting that in 3 of the six galaxies (NGC 4790, NGC 4592 and NGC 7162), we do not find evidence for gas at very high dispersions. It is possible that such a component is present but that our data do not have the spectral resolution or sensitivty to identify this. NGC 1084 shows higher dispersions in the second component in regions with little star formation. NGC 3521 shows a biconical structure in the second component, which resembles an outflow. The velocity maps of NGC 3521 and NGC 4030 clearly show lower rotation velocities for the second component and also a less pinched velocity field.

\begin{figure*}
  \includegraphics[width=\textwidth]{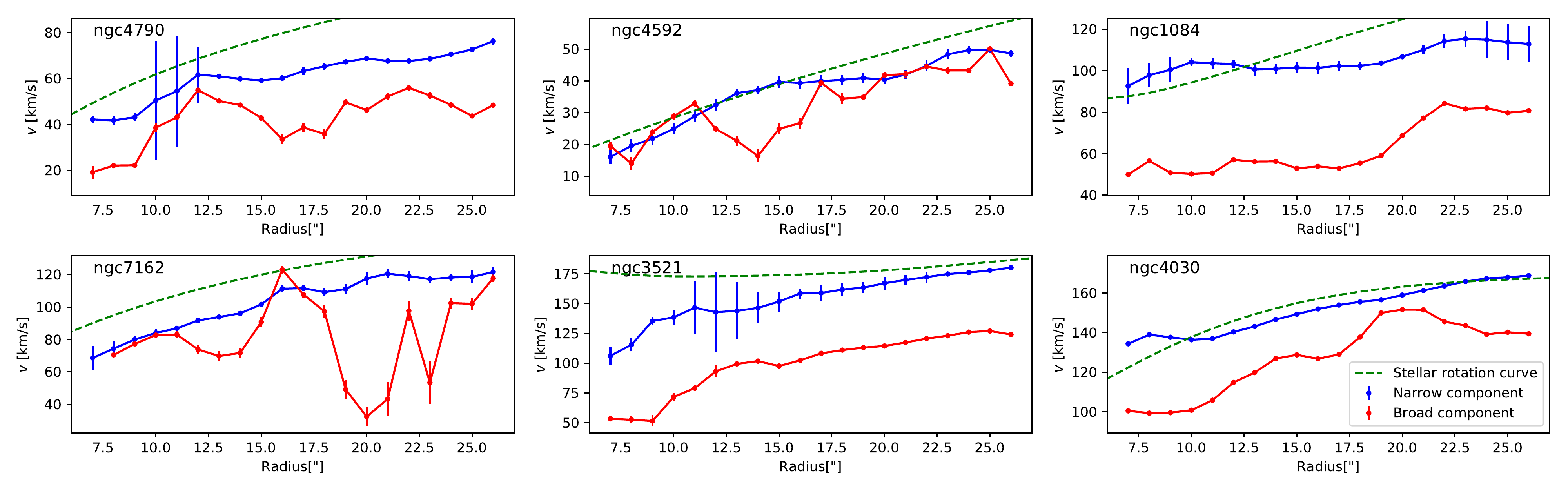}\\
  \vspace{0.5cm}
\includegraphics[width=\textwidth]{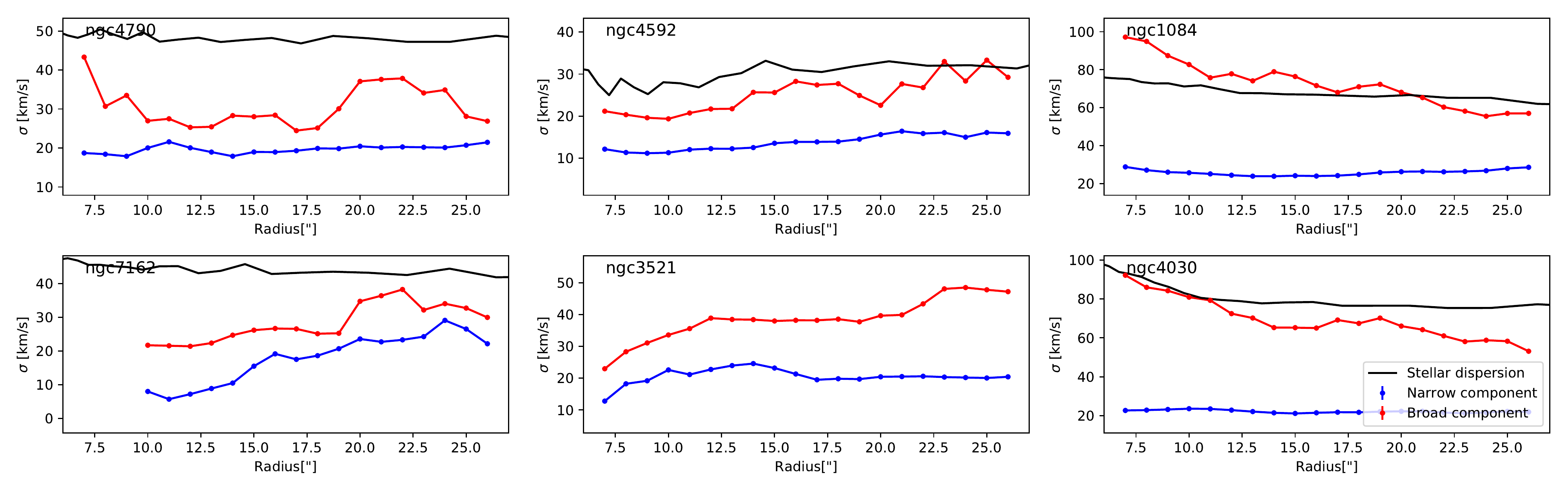}
\caption{Radial profiles of rotation velocity (upper panels) and velocity dispersion (lower panels) for the broad (red) and narrow (blue) components in the six galaxies for which we fitted the kinematics with two Gaussian components. For comparison we show the stellar rotation curve in the upper panels in green, and the stellar velocity dispersions in black. The narrow component have on average low sigma and rotate close to the rotation curve velocity of the midplane. The broad component rotates slower, in accordance with expectiations of the DIG coming from different spatial locations. The dispersions of the broad components are significantly higher than the dispersions of the narrow component, but on average do not exceed the dispersions of the stars. This is consistent with the emission coming from a layer that is thinner than the stellar disk.}\label{fig:2dcomp_vel_dig_f0_ext_1}
\end{figure*}

In Fig. \ref{fig:2dcomp_vel_dig_f0_ext_1} we show a quantitative comparison between the velocities and velocity dispersions of the narrow and broad components and the stellar rotation curves and stellar velocity dispersions. We derive the profiles by fitting the mean radial velocity/velocity dispersion profile of each component. As in particular for the broad component there are many outliers, we clip all values which are more than 5$\sigma$ (standard error) away from the mean value. We then fit the values again (i.e. we do this only for one iteration).
The narrow component is rotating with a velocity close to the circular velocity in the midplane, while the broad component is rotating slower. The dispersion values of the broad component are (per definition) higher than the narrow component, but, except for NGC 1084, do not seem to exceed the values found for the stars. Although differences between SF gas and DIG are more pronounced, these tentative results strengthen our conclusion that majority of the DIG that we identify in our data is coming from gas close to the midplane. 

We additionally check for the presence of high-dispersion gas by stacking
  the continuum subtracted cubes. We use linear interpolation to convert the
  Voronoi binned spectra (of Sec. \ref{sec:kin}) to rest-frame wavelength
  using the \halpha\ velocity. We then carry out a similar two-component fit
  to the stacked spectrum of each galaxy. For the second component, we find
  dispersion values as summarized in Table \ref{tab:apx_disp}. We note that
  these values are slightly higher than those found from the radial fits in
  Fig. \ref{fig:2dcomp_vel_dig_f0_ext_1}. As the lags in velocity between
  the two component fits are sometimes of order 50 km/s, the stacking will
  artificially broaden either component by the same order of
  magnitude. Another reason for the discrepancy is the inclusion of the
  central 7 arc seconds in the data when we stack the cube. When we exclude 
  these central 7 arc seconds we find values that are only slightly higher
  than the values in Fig. \ref{fig:2dcomp_vel_dig_f0_ext_1}. Fits with 3 kinematic components did not converge.

\begin{table}
\caption{Dispersion measurements of the broad component in stacked spectra. We
present measurements for the stacking of the full cube as well as for
excluding the central 7 arcsec.}
\label{tab:apx_disp}  
\begin{tabular}{lcc}
\hline
Name  & $\sigma$ full cube & $\sigma$ centre excluded\\
      &    [km/s]          &  [km/s] \\
\hline
NGC4790  & $33.5_{-0.6}^{+0.7}$  &  $54.9_{-15}^{+110}$    \\
NGC4592 & $31.3_{-0.5}^{+0.7}$  &  $29.4_{-7}^{+17}$       \\
NGC1084  & $180_{-80}^{+5}$  &  $136_{-35}^{+68}$     \\
NGC7162 & $163_{-80}^{+151}$  &  $85_{-38}^{+195}$      \\
NGC3521  & $126_{-34}^{+2}$  &  $59.6_{-9}^{+17}$      \\
NGC4030  & $143_{-3}^{+2}$  &  $103_{-29}^{+37}$     \\
\end{tabular}
\end{table}

\section{Stellar rotation curves}\label{sec:rotcurve}
We derive an estimate of the circular velocity in the midplane of the subsample of 6 galaxies by fitting Jeans Anisotropic Models (JAM) by \citet{Cap08}  to the second moment of stellar kinematics. This second moment was derived by using $v^2_{RMS} = \sigma^2 + V^2_{los}$. The free parameters of these models are a multi Gaussian expansion (MGE) of the mass distribution in the galaxy, an MGE expansion of the light distribution, a value for the orbital anisotropy $\beta_z$ and the inclination.

We fix the inclination to the value used in this paper based on the flattening of the galaxy light. We use the structural parameter fits of \citet{SalLauLai15}  to derive the luminous distribution of the tracer population, by converting each morphological component into an MGE using the \texttt{mge\_fit\_1d} code \citep{Cap14}. These different luminous components were then scaled by a M/L ratio (per morphological component) and combined to form the MGE of the mass distribution. We allowed  $\beta_z$ to vary between -1.0 and 0.5. The JAM models allow for a convolution with a PSF, for which we used a single Gaussian with a FWHM of 0\farcs7, which is typical for the sample.

We  use the best-fit JAM models to obtain a mass distribution of the galaxy. Using the same deprojection as in the JAM models, we calculate the circular velocity in the $z=0$ plane  and multiply this by $\sin(i)$ to correct for the inclination. Given the crude modeling (e.g. the assumptions on the light profile, no proper dark matter halo), this velocity is more indicative than accurate.

\section{Results for constrained fits to $f_0$}\label{apx:newf0}
In the text we followed Blanc et al. to determine the contribution of emission of the DIG to the observed \halpha\ surface brightness. We assumed that there was a value, $f_0$, below which the surface brightness of \halpha\ is completely dominated by the DIG. We determined a galaxy wide value of $f_0$ by fitting the observed line ratios of  [\ion{S}{2}]/\halpha\ as a function of \halpha\ surface brightness, assuming a free scaling, assumed to be a metallicity, between the intrinsic values of  [\ion{S}{2}]/\halpha for DIG and \ion{H}{2} regions in the Milky Way, and those in the galaxy we fitted. Here we perform a similar fit, but instead do not use a free parameter for the metallicity. We use instead the values determined in Paper 2 with the M13 method and assume a solar abundance of $12+\log$ O/H = 8.69.

\begin{figure*}
  \includegraphics[width=\textwidth]{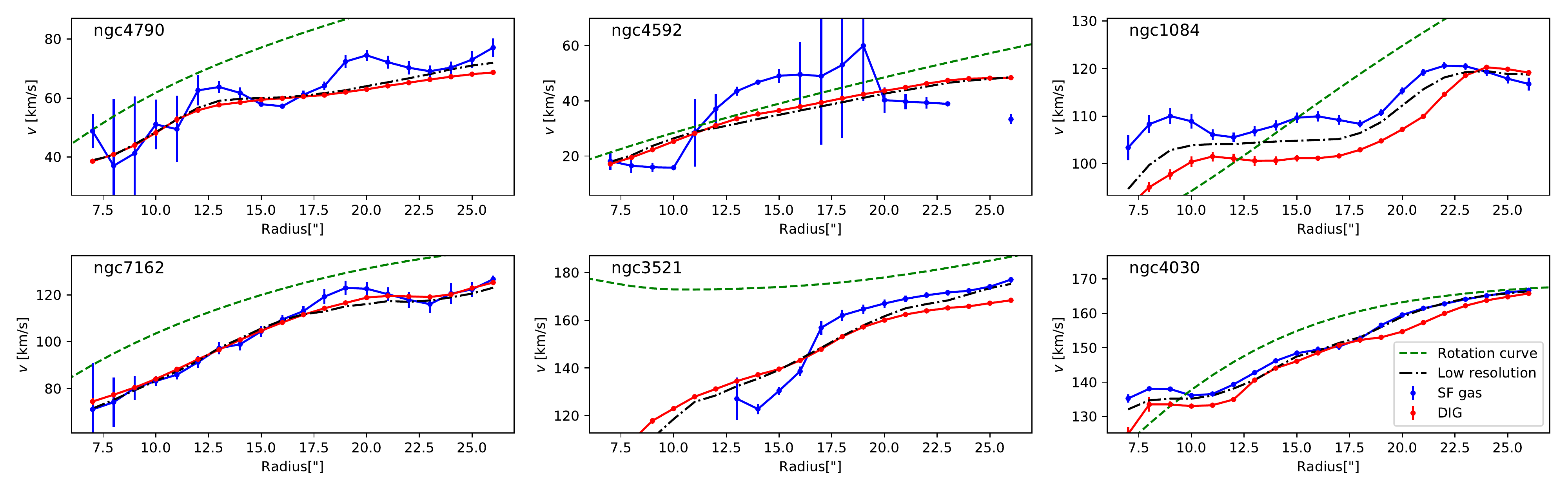}\\
 \caption{Rotation velocities of star forming and diffuse gas in a subset of the 6 most regularly rotating galaxies in our sample. Same figure as Fig. \ref{fig:vel_dig_f0_ext}, except that we determined the differences in velocity with metallicity-constrained values of $f_0$.}
 \label{fig:vel_dig_f0_ext_newf0}
\end{figure*}

\begin{figure}
   \includegraphics[width=0.45\textwidth]{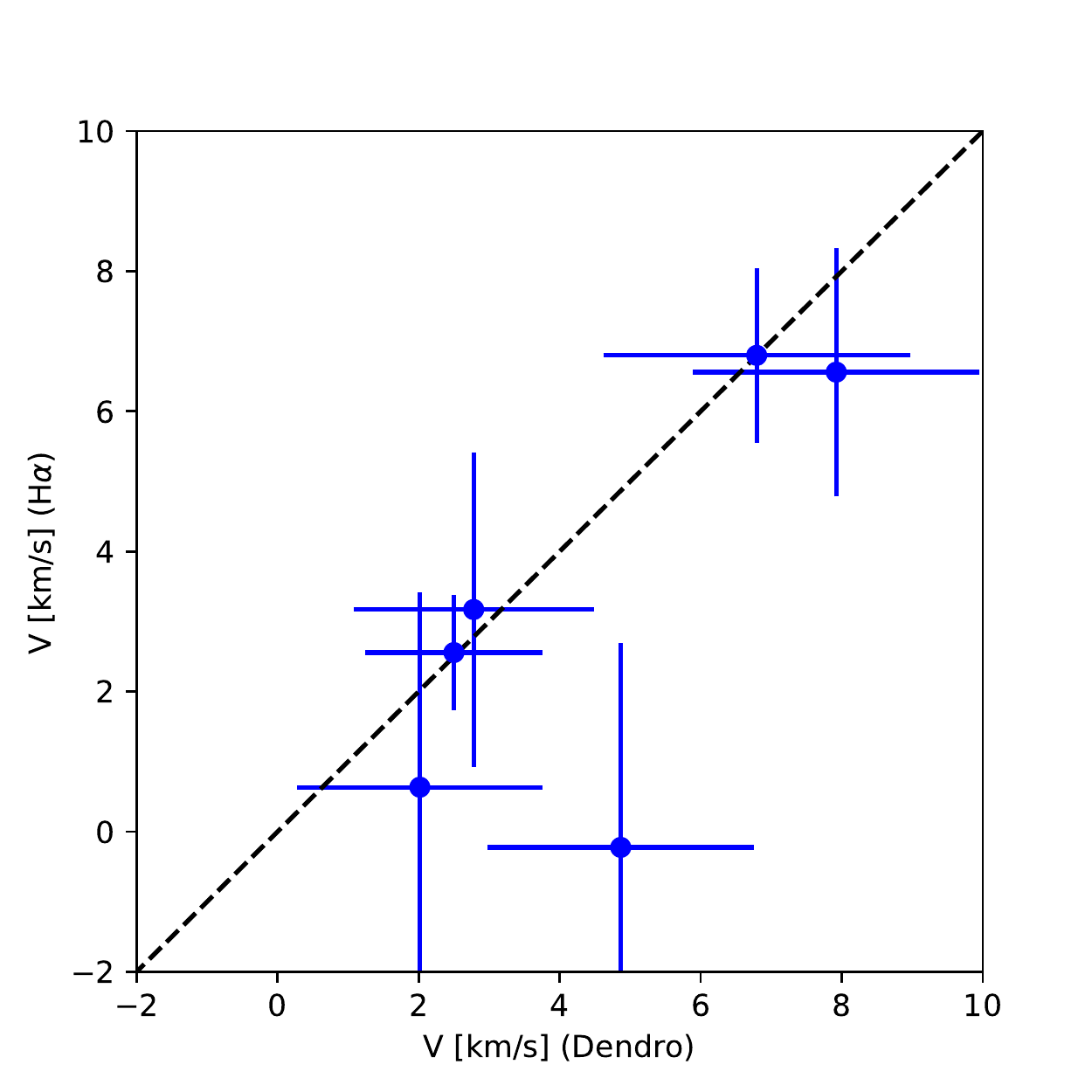}
 \caption{
Median velocity difference between SF gas and DIG for the Dendrogram method and the metallicity constrained $f_0$ method. }
 \label{fig:prop_newf0}
\end{figure}

The $f_0$ values that we find for these more constrained fits are higher than the values found in the text using metallicity as a free parameter. The fits to the [\ion{S}{2}]/\halpha\ line ratios with the metallicity fixed to the Paper 2 value, do not look as good as those with one more free parameter used in the main text of this Paper. For completeness, we do include the results of a repeated analysis with these somewhat higher values for $f_0$ in Figures \ref{fig:vel_dig_f0_ext_newf0} and \ref{fig:prop_newf0}.

%\clearpage
\section{Identification of DIG}\label{apx:id_plots}
\begin{figure*}
 \includegraphics[width=0.8\textwidth]{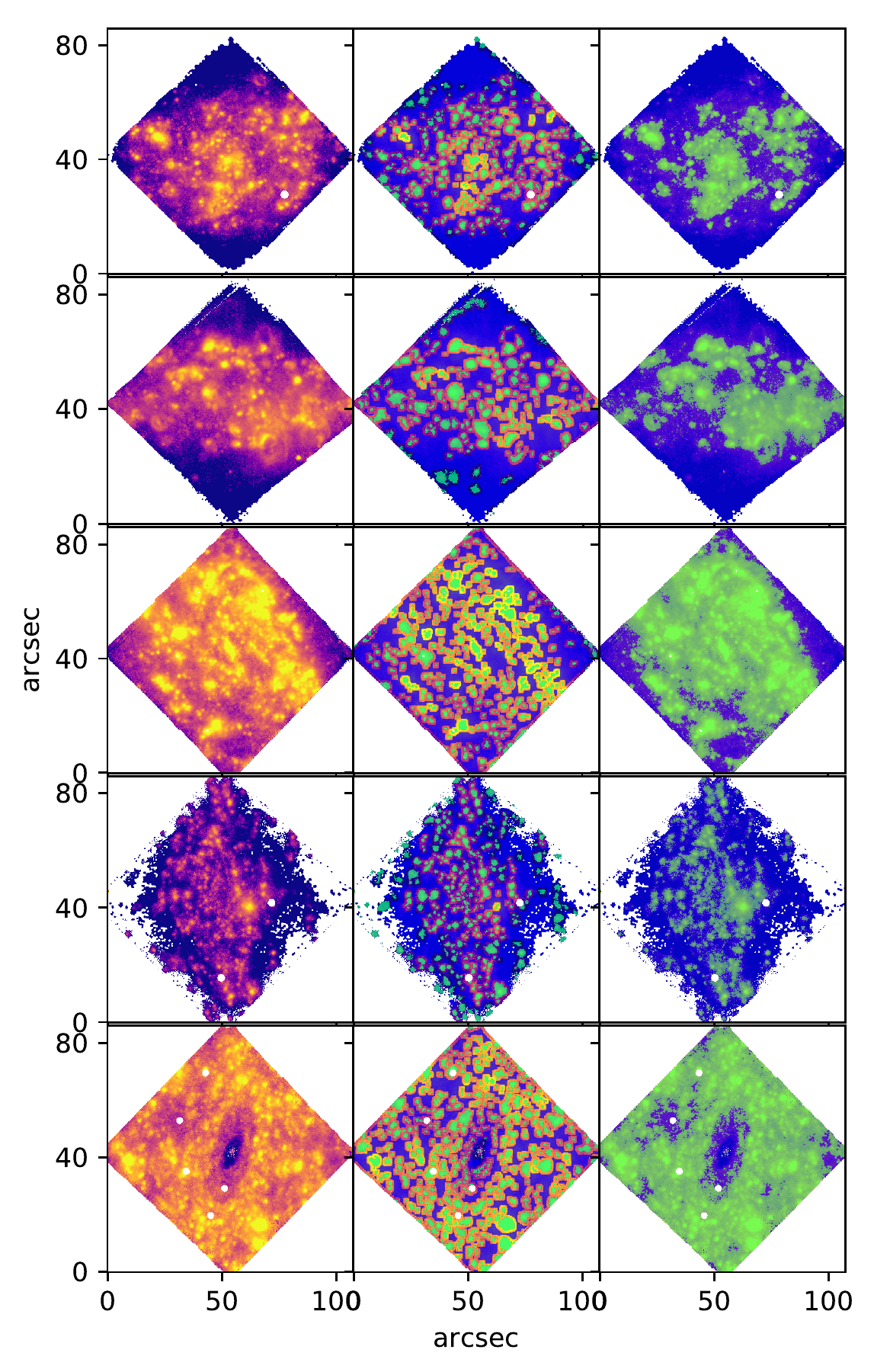}
 \caption{Same as Fig. \ref{fig:ngc4030_dig_id}, but for the galaxies NGC
   4790, NGC 4592, NGC 1084, NGC 7162 and NGC 3521}
 \label{fig:dig_id_all}
\end{figure*}

\begin{figure*}
  \includegraphics[width=0.99\textwidth]{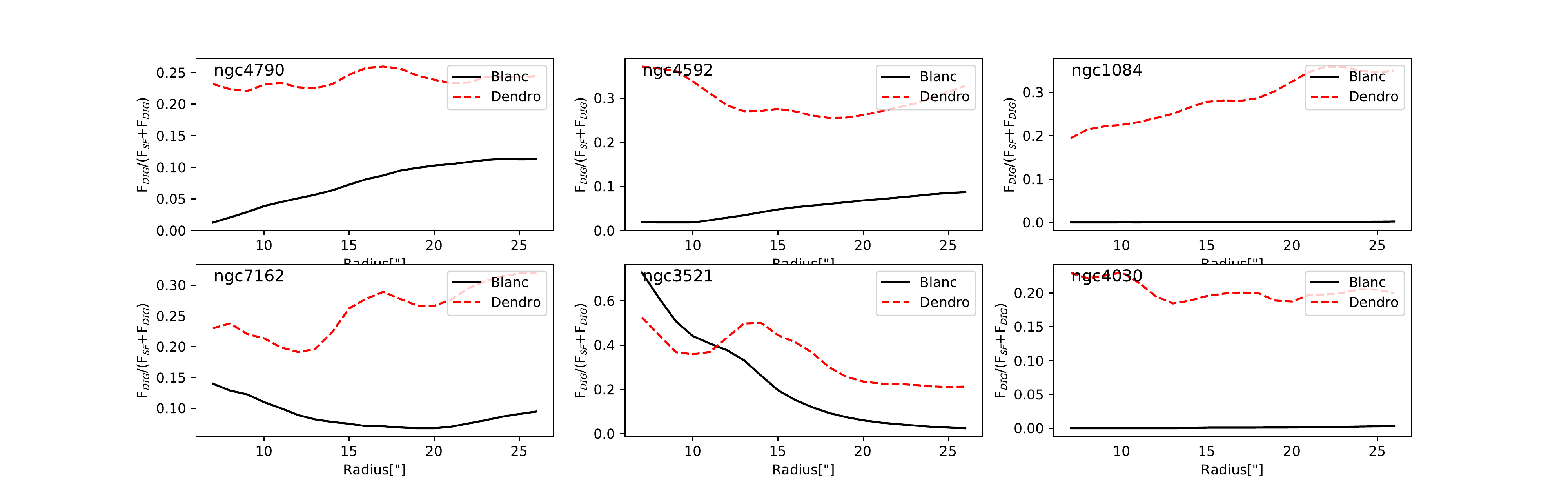}
  \includegraphics[width=0.99\textwidth]{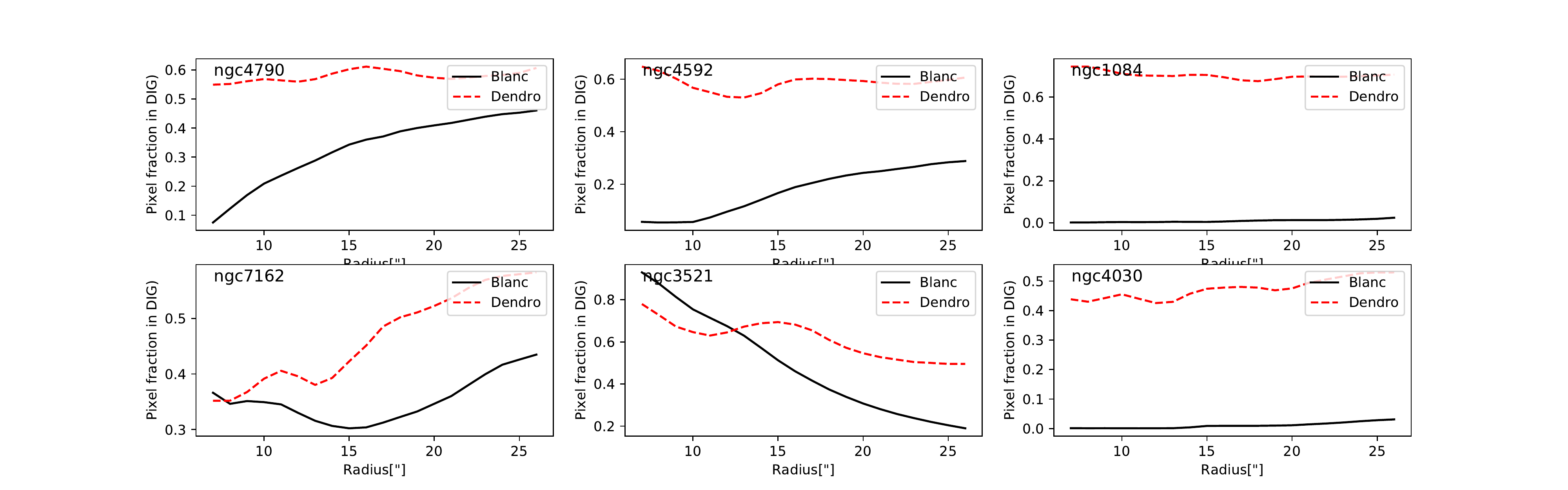}
 \caption{The fraction of light and pixels in the SF and DIG components of
   each galaxy. The fraction of DIG as identified by the dendrogram method is
   significantly higher than using the Blanc criterion.}
 \label{fig:dig_frac}
\end{figure*}

{
In Fig. \ref{fig:dig_id_all} in this Appendix we show the the identification of DIG superimposed on
\halpha\ maps. For comparison with lower resolution studies that lack spatial
resolution to distinguish between DIG and SF gas, we show in Figure
\ref{fig:dig_frac} the light fraction and pixel fraction of SF gas and DIG.}

%%%%%%%%%%%%%%%%%%%%%%%%%%%%%%%%%%%%%%%%%%%%%%%%%%

% Don't change these lines
\bsp	% typesetting comment
\label{lastpage}
\end{document}